\documentclass[pre,aps,amsmath,amssymb,reprint,groupedaddress]{revtex4-2} 

\usepackage[T1]{fontenc}
\usepackage[utf8]{inputenc}
\usepackage{lmodern}
\usepackage{pgf}
\usepackage{tikz}
\usepackage{booktabs}
\usepackage{scalerel}
\DeclareUnicodeCharacter{2212}{\ensuremath{-}}

\usetikzlibrary{svg.path}
\definecolor{orcidlogocol}{HTML}{A6CE39}
\tikzset{
  orcidlogo/.pic={
    \fill[orcidlogocol] svg{M256,128c0,70.7-57.3,128-128,128C57.3,256,0,198.7,0,128C0,57.3,57.3,0,128,0C198.7,0,256,57.3,256,128z}; 
    \fill[white] svg{M86.3,186.2H70.9V79.1h15.4v48.4V186.2z}
                 svg{M108.9,79.1h41.6c39.6,0,57,28.3,57,53.6c0,27.5-21.5,53.6-56.8,53.6h-41.8V79.1z M124.3,172.4h24.5c34.9,0,42.9-26.5,42.9-39.7c0-21.5-13.7-39.7-43.7-39.7h-23.7V172.4z} 
                 svg{M88.7,56.8c0,5.5-4.5,10.1-10.1,10.1c-5.6,0-10.1-4.6-10.1-10.1c0-5.6,4.5-10.1,10.1-10.1C84.2,46.7,88.7,51.3,88.7,56.8z}; 
  }
}

\newcommand\orcidicon[1]{\href{https://orcid.org/#1}{\mbox{\scalerel*{
\begin{tikzpicture}[yscale=-1,transform shape]
\pic{orcidlogo};
\end{tikzpicture}
}{D}}}}

\usepackage[citecolor=blue,colorlinks=true,pdfusetitle,
pdfauthor={Vishnu V. Krishnan, Smarajit Karmakar, Kabir Ramola},
pdfsubject={Statistical Physics of Disordered systems},
pdfkeywords={Glass, Hessian, Asymptotes, Poles, Singularity}]{hyperref}

\begin{document}
\title{Singularities in Hessian element distributions of amorphous media}

\author{Vishnu \surname{V. Krishnan}\thinspace\orcidicon{0000-0003-3889-3214}} 
\email{vishnuvk@tifrh.res.in}
\author{Smarajit Karmakar\thinspace\orcidicon{0000-0002-5653-6328}} 
\email{smarajit@tifrh.res.in}
\author{Kabir Ramola\thinspace\orcidicon{0000-0003-2299-6219}} 
\email{kramola@tifrh.res.in}
\affiliation{Centre for Interdisciplinary Sciences, Tata Institute of
Fundamental Research, Hyderabad 500046, India}

\date{\today}

\begin{abstract}
    We show that the distribution of elements $H$ in the Hessian matrices
    associated with amorphous materials exhibit singularities $P(H) \sim
    {\lvert H \rvert}^{\gamma}$ with an exponent $\gamma < 0$, as $\lvert H
    \rvert \to 0$. We exploit the rotational invariance of the underlying
    disorder in amorphous structures to derive these exponents exactly for
    systems interacting via radially symmetric potentials. We show that
    $\gamma$ depends only on the degree of smoothness $n$ of the potential of
    interaction between the constituent particles at the cut-off distance,
    independent of the details of interaction in both two and three dimensions.
    We verify our predictions with numerical simulations of models of
    structural glass formers.  Finally, we show that such singularities affect
    the stability of amorphous solids, through the distributions of the minimum
    eigenvalue of the Hessian matrix.
\end{abstract}


\maketitle


\textit{Introduction:}
Understanding and modelling the properties of amorphous solids such as glasses
has remained a challenge due to their extreme non-equilibrium nature as well as
the underlying disorder in the arrangement of
particles~\cite{angell1988perspective, kirkpatrick1987p,
*kirkpatrick1988comparison, *kirkpatrick1988mean, *kirkpatrick1989random,
debenedetti1996metastable, das2004mode, gotze2008complex,
shintani2008universal, berthier2011theoretical, wolynes2012structural,
karmakar2014growing, karmakar2015length}. An important aspect in the study of
such systems is their vibrational properties, typically probed through the
eigenvalues of the Hessian matrix~\cite{sastry1997statistical,
angell2003hyperquenched, zaccone2020relaxation}, with low-frequency modes being
particularly relevant at the low-temperatures where glass physics
dominates~\cite{angell1972configurational, goldstein1976viscous,
buchenau1984neutron, ruffle2003observation, gurevich2003anharmonicity,
ruffle2006glass, moriel2019wave, monaco2009anomalous, monaco2009breakdown,
degiuli2014pressure, charbonneau2016nondebye, lerner2016vibration,
kapteijns2018nonphononic, wang2019low, zaccone2020relaxation, shimada2020low}.
Since glass forming systems settle into disordered configurations, random
matrix treatments provide a natural framework with which to model amorphous
systems~\cite{stanifer2018random, beltukov2011sparse, manning2015peak,
beltukov2013ioffe, baggioli2019vibrational, conyuh2017bosonpeak,
brito2010granular, altieri2019, maradudin1958disordered, gurarie2003bosonic,
benetti2018mean}.  The Hessian is thus naturally characterised by the
distribution of its elements~\cite{cavagna1999analytic, huang2009localization}.
Although the Hessian matrix is relatively simple to compute in
simulations~\cite{kittel1976introduction, ashcroft1976solid}, an experimental
determination remains difficult, accessible for example, only through
displacement correlations in colloidal glasses~\cite{ghosh2010density}.  In
this context, it is important to study the vibrational properties of model
systems via simulations. Simulation models vary in strength and range of
interaction~\cite{xipeng2020potential}, and exhibit a wide range of physical
properties~\cite{dauchot2011brittle}.  Even though there are indications of an
underlying ``amorphous order'' in such systems~\cite{karmakar2014growing,
karmakar2015length}, their behaviour differs fundamentally from that of
crystals, with the low lying eigenvalues of their Hessian matrices displaying
marked non-Debye behaviour~\cite{monaco2009anomalous, monaco2009breakdown,
degiuli2014pressure, charbonneau2016nondebye, lerner2016vibration,
kapteijns2018nonphononic, wang2019low, shimada2020low}, related to the growth
of a structural length scale~\cite{karmakar2012direct} as well as static
correlations~\cite{rainone2020pinching}. The distribution of elements affects
the eigenvalues of the Hessian which is directly related to the mechanical
properties of amorphous solids~\cite{maloney2004failure, maloney2006shear,
dasgupta2012instability}.

\begin{figure}[t]
    \resizebox{0.95\linewidth}{!}{\input{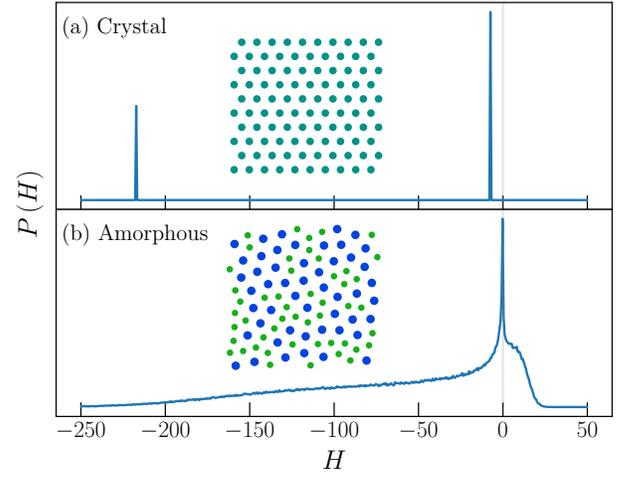}}
    \caption{Comparison of energy minimised configurations in (a) a
    mono-disperse crystal and (b) an amorphous glass consisting of two types of
    particles: A (small, green) and B (large, blue). Also plotted are the
    corresponding distributions $P(H)$ of the diagonal elements of their
    Hessians. In (b), the distribution is shown only for A-B interactions.}
    \label{fig_CrysAmore_2dR10} 
\end{figure}

In this Letter, we analyse the distributions of Hessian elements in structural
glass formers analytically, as well as through numerical simulations.  In
simulations, interaction potentials are typically cut-off at a finite distance
for computational expediency, and in order to avoid unphysical changes, they
are smoothed to relevant degrees at this
cut-off~\cite{frenkel2001understanding}. However, the effect of this smoothness
on vibrational properties of such amorphous systems has never been
investigated.  In crystals, the ordered inter-particle distances imply delta
distributed Hessian elements.  For example, in a triangular lattice of
particles with nearest neighbour interactions as shown in
Fig.~\ref{fig_CrysAmore_2dR10}~(a), the diagonal Hessian elements take on one
of two non-zero values. However, in the case of a disordered arrangement
derived from simulations of glass formers as shown in
Fig.~\ref{fig_CrysAmore_2dR10}~(b), this distribution is continuous, peaking at
zero with a marked singularity. We exploit the rotational invariance of
the underlying amorphous disorder to derive analytic expressions for these
distributions, for systems with radially symmetric interactions, which we then
verify using direct numerical simulations. We show that these distributions
indeed exhibit smoothness-dependent singularities
\begin{equation}
    P\left( H \right) \sim {\left\lvert H
    \right\rvert}^{\gamma}, ~~~\text{for}~~~ |H| \rightarrow 0
    \label{eq_singularity}
\end{equation}
with a power $\gamma < 0$. We derive exact results for $\gamma$, for any degree
of smoothness, in both two dimensions (2D) as well as three dimensions (3D).
Our results are summarized in Table~\ref{tab_results}, highlighting the
non-trivial dependence of these distributions on the nature of the interaction
at the cut-off.  Finally, we show that these singularities have crucial
implications for the low-energy vibrational modes of amorphous systems, through
a numerical sampling of the minimum eigenvalue of systems with varying
smoothness in their interactions.

\begin{figure}[t!]
    \resizebox{0.95\linewidth}{!}{\input{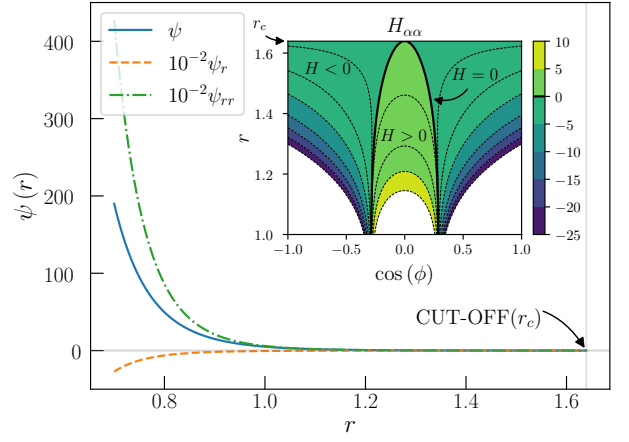}}
    \caption{The R10 potential $\psi(r) \sim r^{-10} + c + b r^{2}+ a r^{4}$
    for A-B interactions, smoothed to two derivatives $(n = 2)$ at the cut-off
    $[\psi(r_{c}) = \psi^{\prime}(r_{c}) = \psi^{\prime \prime}(r_{c}) = 0]$.
    The derivatives have been scaled down by a factor of 100 for clarity. The
    inset shows a plot in the inter-particle coordinates $(r,\cos \phi)$,
    displaying contours which contribute to the distribution of diagonal
    Hessian elements ($\alpha = \beta$) at a fixed $H$.}\label{fig_countour}
\end{figure}

\begin{figure}[t!]
    \resizebox{0.95\linewidth}{!}{\input{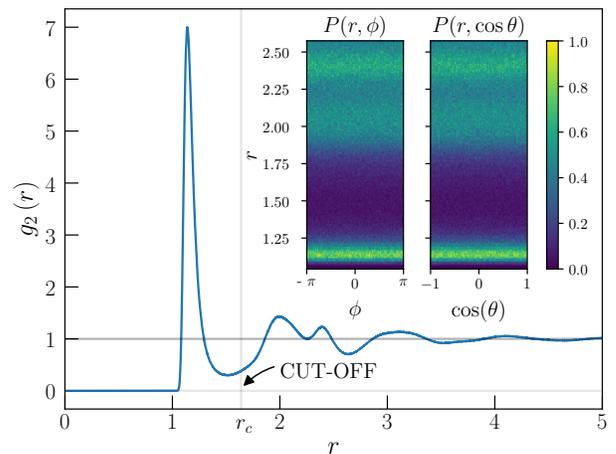}}
    \caption{Numerically sampled radial distribution function of A-B particle
    pairs in the R10 system in three dimensions. The vertical line indicates
    the cut-off distance of the interaction. Inset: Numerically sampled joint
    distributions of the distances ($r$) and angles ($\phi$ and $\cos\theta$),
    indicating isotropic angular distributions.}\label{fig_isotropy}
\end{figure}

\textit{Distribution of Hessian Elements:}
We begin with a disordered arrangement of particles at positions
$\{\mathbf{r}^i \}$, where the particle index $i \in \{1, \ldots , N\}$. The
Hessian is then a $d N \times d N$ matrix where $d$ represents the dimension of
the system, with elements $\mathcal{H}^{i j}_{\alpha \beta}$ that describe the
stiffness between particles $i$ and $j$, along the coordinates $\alpha, \beta
\in \{x,y,z\}$.  In terms of the inter-particle distance vector $\mathbf{r}^{i
j} = \mathbf{r}^{i} - \mathbf{r}^{j}$, the Hessian elements may be expressed as
\begin{equation}
    \mathcal{H}_{\alpha \beta}^{i j} (\mathbf{r}^{i j}) = \frac{\partial^{2} U
    \left[\{\mathbf{r}^{i}\}\right]}{\partial r^{i j}_{\alpha} \partial r^{i
    j}_{\beta}}.
    \label{eq_hessian}
\end{equation}
Here, $U \left[\{\mathbf{r}^{i}\}\right]$ is the total potential energy of the
system, which is a function of the positions of all particles. For pairwise
additive interactions, $U\left[\{\mathbf{r}^{i j}\}\right] = \sum_{i j} \psi^{i
j}$, where $\psi^{i j}$ is the interaction potential between particles $i$ and
$j$. We consider central potentials $ \psi(r)$, where $r \equiv r^{i j} =
\lvert \mathbf{r}^{i j} \rvert$ is the distance between particles $i \text{ and
} j$.  In addition, these potentials are smoothed to $n$ derivatives at a
cut-off distance $r_c$, i.e., $\left.  \frac{d^m \psi}{d r^m}
\right\rvert_{r_{c}} = 0$ for all $0 \le m \le n$.  Fig.~\ref{fig_countour}
shows a typical interaction potential in 2D used in our numerical simulations,
and its derivatives, all of which tend to zero at cut-off ($n = 2$ in this
case).  The Hessian elements for central potentials are given
by~\cite{karmakar2010athermal}
\begin{equation}
    \mathcal{H}_{\alpha \beta}^{i j} (\mathbf{r}^{i j}) = - \left(
    \frac{\psi_{r r}^{i j}}{{(r^{i j})}^{2}} - \frac{\psi_{r}^{i j}}{{(r^{i
    j})}^{3}} \right) r^{i j}_{\alpha} r^{i j}_{\beta} - \delta_{\alpha \beta}
    \frac{\psi_{r}^{i j}}{r^{i j}},
    \label{eq_hess}
\end{equation}
where the subscripts of $r$ indicate partial derivatives with respect to the
inter-particle distances: $\psi_{r} \equiv \psi_{r}^{i j}  = \partial_{r^{i j}}
\psi^{i j}, \;  \psi_{r r} \equiv \psi_{r r}^{i j} = \partial_{r^{i j}}
\partial_{r^{i j}} \psi^{i j}$. Since the potential and its derivatives vanish
at the cut-off distance, the small Hessian elements arise primarily due to
pair-distances near this cut-off.  The values of Hessian elements in
Eq.~(\ref{eq_hess}) depend on the length and angle of the inter-particle
distances in an arbitrarily chosen (fixed) Cartesian coordinate system.  We
therefore define a generalised angular coordinate $\Omega$ with respect to
these fixed axes, which is a function of the angle $\phi$ in 2D, and both the
polar and azimuthal angles $\theta, \phi$ in 3D.  The distribution of Hessian
elements $P(H)$ is then given by
\begin{equation}
    P \left( H \right) = \int d r \; d \Omega \; P \left( r, \Omega \right)
    \delta{\left( H - \mathcal{H}_{\alpha \beta}^{i j} (\mathbf{r}^{i
    j})\right)},\label{eq_probhess}
\end{equation}
where $P(r,\Omega)$ represents the joint probability distributions of
inter-particle distances and orientations, as shown in the inset of
Fig.~\ref{fig_isotropy}.  For a given $H$, the delta function constraint in
Eq.~(\ref{eq_probhess}) selects the corresponding contour in $(r,\Omega)$
through Eq.~(\ref{eq_hess}), as shown in the inset of Fig.~\ref{fig_countour}.

We next exploit the rotational invariance of the underlying amorphous disorder
to relate $P(H)$ to the distribution of inter-particle distances. This
microscopic rotational symmetry leads to isotropic angular distributions at all
distances. Hence, we may assume the distributions of inter-particle distances
and angles to be uncorrelated $P(r,\Omega) = P(r) P(\Omega)$. Numerical
sampling of this joint distribution $P(r,\Omega)$ in structural glass formers
(inset of Fig.~\ref{fig_isotropy}), demonstrates that all orientations are
sampled uniformly, independent of the radial distance. We therefore assume
uniform distributions in the angular variables $P_{2\text{D}} (\phi) =
P_{3\text{D}} (\phi) = \frac{1}{2 \pi}$, and $P_{3\text{D}} (\cos \theta) =
\frac{1}{2}$.  The distribution of $r$ is the radial distribution function
$g_2(r)$, normalised over the range of interaction $r \in [0,r_{c}]$, i.e.,
$P(r) = g_{2}(r) \Theta(r_{c} - r) \mathcal{N}$, where $\Theta$ is the
Heaviside function, and ${\mathcal{N}}^{-1} = \int_{0}^{r_{c}} g_{2}(r)$ is a
constant of normalisation.  For a given $H$, the integration over the angular
variables in Eq.~(\ref{eq_probhess}) yields a Jacobian factor, leading to the
Hessian element distribution
\begin{equation}
    P \left( H \right) = \int_{0}^{r_c} d r \; P(r) \; \frac{P \left( \Omega
    \right)}{\left\lvert \frac{\partial H}{\partial \Omega} \right\rvert} =
     \int_{0}^{r_c} d r \; P(r) \; \mathcal{P} (H, r).
    \label{eq_probhess_partint}
\end{equation}
Above, the angular variable $\Omega$ in the final expression has been expressed
in terms of $H$ and $r$ using Eq.~(\ref{eq_hess}).  Therefore, given an
empirical $g_2(r)$, $P(H)$ can be determined exactly.  The partial integrand
$\mathcal{P}(H,r) = {\left[ \left\lvert \frac{\partial H}{\partial \Omega}
\right\rvert \right]}^{-1} P \left( \Omega \right)$ depends on the dimension,
as well as the spatial indices $\alpha$ and $\beta$.  However, rotational
symmetry dictates that there are only two classes of Hessian element
distributions: the diagonal ($H_{\alpha \alpha} : H_{x x} \equiv H_{y y} \equiv
H_{z z}$) and off-diagonal ($H_{\alpha \beta} :  H_{x y} \equiv H_{y z} \equiv
H_{z x}$).

\textit{Exact Partial Integrands:}
We first consider the distribution of Hessian elements in 2D systems, choosing
the diagonal element $H_{x x}$ and the off-diagonal element $H_{x y}$.  From
Eq.~(\ref{eq_hess}), $ H^{2\text{D}}_{x x} = - \left( \psi_{r r} -
\frac{\psi_{r}}{r} \right) \cos^{2} \phi - \frac{\psi_{r}}{r}$, and
$H^{2\text{D}}_{x y} = - \left( \psi_{r r} - \frac{\psi_{r}}{r} \right) \cos
\phi \sin \phi$.  It is convenient to use the angular variable $\Omega \equiv
\cos \phi$ with the distribution $P_{2\text{D}} \left( \Omega \right) = 1 /
\left(\pi \sqrt{1 - \Omega^{2}}\right)$.  Substitution of these into the
expression in Eq.~(\ref{eq_probhess_partint}) yields the same partial integrand
for both cases $\mathcal{P}^{2\text{D}}_{\alpha \beta} = {\left[ \pi
\left\lvert \frac{\partial H}{\partial \cos \phi} \right\rvert \sqrt{1 -
\cos^{2} \phi} \right]}^{-1}$.  Finally, inverting the above expressions for $
H^{2\text{D}}_{x x}$ and $H^{2\text{D}}_{x y}$ to express $\cos \phi$ in terms
of $H$ and $r$, we arrive at
\begin{align}
    \mathcal{P}^{2\text{D}}_{\alpha \beta} \left( H, r \right) =
    \begin{cases}
        {\left\lvert 4 \pi^{2} \left( H + \frac{\psi_{r}}{r} \right) \left( H +
        \psi_{r r} \right) \right\rvert}^{-1/2} & \alpha = \beta, \\
        {\left\lvert \pi^{2} \left\{ {\left( \psi_{r r} - \frac{\psi_{r}}{r}
        \right)}^{2} - 4 H^{2} \right\} \right\rvert}^{-1/2} & \alpha \neq
        \beta.
    \end{cases}
    \label{eq_partint_2D}
\end{align}
For 3D systems, we choose $H_{z z}$ and $H_{x z}$ to represent the diagonal and
off-diagonal elements, respectively. From Eq.  (\ref{eq_hess}) we have $
H^{3\text{D}}_{z z} = - \left( \psi_{r r} - \frac{\psi_{r}}{r} \right) \cos^{2}
\phi -\frac{\psi_{r}}{r}$, and  $H^{3\text{D}}_{x z} = - \left( \psi_{r r} -
\frac{\psi_{r}}{r} \right) \cos \theta \sin \theta \cos \phi$.  In the case of
diagonal elements, it is convenient to choose $\Omega \equiv \cos \theta$, with
the distribution $P_{3\text{D}}  \left( \cos \theta \right) = \frac{1}{2}$ (see
inset of Fig.~\ref{fig_isotropy}).  Under these conditions, the partial
integrand defined in Eq.~(\ref{eq_probhess_partint}) is
$\mathcal{P}^{3\text{D}}_{z z} = {\left[ 2 \left\lvert \frac{\partial
H}{\partial \cos \theta} \right\rvert \right]}^{-1}$. In the case of
off-diagonal elements, it is convenient to choose $\Omega \equiv \cos \phi$,
and given that $P  \left( \phi \right) = \frac{1}{2}$, we arrive at the exact
integral form $\mathcal{P}^{3\text{D}}_{x z} = \int \pi d (\cos \theta) {\left[
    \left\lvert \frac{\partial H}{\partial \cos \phi} \right\rvert \sqrt{1 -
    \cos^{2} \phi} \right]}^{-1}$.  Finally, inverting the above expressions
for $ H^{3\text{D}}_{z z} \text{ and } H^{3\text{D}}_{x z}$ to express $\cos
\theta \text{ and } \cos \phi$ in terms of $H$ and $r$, we arrive at the
simplified expressions
\begin{align}
    \mathcal{P}^{3\text{D}}_{\alpha \beta} \left( H, r \right) =
    \begin{cases}
        {\left\lvert 4 \left( H + \frac{\psi_{r}}{r} \right) \left(\psi_{r r} -
        \frac{\psi_{r}}{r} \right) \right\rvert}^{-1/2} & \alpha = \beta, \\
        \frac{\kappa}{H} \int^{1}_{-1} dx {\left[ x^{2} (1 - x^{2}) -
        \kappa^{2} \right]}^{-1/2} & \alpha \neq \beta. \\
    \end{cases}
    \label{eq_partint_3D}
\end{align}
with $\kappa  = H {\left( \psi_{r r} - \frac{\psi_{r}}{r} \right)}^{-1}$.  The
above integral form for the off-diagonal elements has the asymptotic behaviour
(refer to Supplemental Material~\cite{SI}/\ref{appendix_exact_part_int})
 \begin{equation}
    \mathcal{P}^{3\text{D}}_{\alpha \neq \beta} \left( H, r \right)
     \overset{\kappa \to 0}{\sim} \frac{2 \kappa}{H} \log{\left(
     \frac{4}{\kappa} \right)}.
    \label{eq_partint_limit_OffDiag_3D}
\end{equation}

The distributions of Hessian elements can now be found using these partial
integrands  [Eqs.~(\ref{eq_partint_2D})~and~(\ref{eq_partint_3D})] in
Eq.~(\ref{eq_probhess_partint}) along with the numerically obtained radial
distribution functions, as shown in Fig.~\ref{fig_isotropy}. We display the
match between our theoretical predictions and distributions obtained from
numerical simulations in Fig.~\ref{fig_hess_distro}.  In the Supplemental
Material~\cite{SI}/\ref{appendix_exact_g2r}, we additionally show a precise
match between $P(H)$ obtained analytically and numerically, for an
exponentially decaying $g_{2}(r)$.

\textit{Asymptotic Forms and Singular Distributions:}
We focus next on the behaviour of $P(H)$ in the limit $H \to 0$.  It is clear
from the forms of the partial integrands in
Eqs.~(\ref{eq_partint_2D})~and~(\ref{eq_partint_3D}), that they diverge as $r
\to r_c$ and $H \to 0$, since $\psi_{r} \to 0 ~\text{ and }~ \psi_{r r} \to 0$.
Consequently, the corresponding integrals determining $P(H)$ in
Eq.~(\ref{eq_probhess_partint}) exhibit singularities $P(H) \sim H^{\gamma}$ as
$H \to 0$, as displayed in Figs.~\ref{fig_CrysAmore_2dR10}~(b)
and~\ref{fig_hess_distro}. We show below that the strength $\gamma$ of this
singularity depends on the rate of decay (i.e., smoothness $n$) of the
interaction at cut-off.

To determine the behaviour of the distribution near the singularity, we focus
on the integral in Eq.~(\ref{eq_probhess_partint}), as we approach the cut-off
$r \to r_c$. It is assumed also, that the $g_{2}(r)$ does not contain
singularities at $r_{c}$ as justified by the plot in Fig.~\ref{fig_isotropy}.
The interaction potential, smooth to $n$ derivatives at the cut-off distance
and its derivatives in the limit of $r \to r_{c}$, may therefore be
approximated as
\begin{align}
    \psi \left(r\right) &= {\left( r_{c} - r \right)}^{n+1} f \left(r\right), \nonumber \\
    \psi_{r} / r &\approx C_{1} {\left( r_{c} - r \right)}^{n}, \nonumber \\
    \psi_{r r} &\approx C_{2} {\left( r_{c} - r \right)}^{n-1}, \label{eq_approx}
\end{align}
where $f(r)$ is a regular function, along with $C_{1} = - (n+1) f(r) / r$, and
$C_{2} = n (n+1) f(r)$ which vary significantly slower compared to the
power-law term as $r \to r_{c}$. Above, in $\psi_{r} / r$, we have ignored
${(r_{c} - r)}^{n+1} f'(r)$ in comparison to ${(r_{c} - r)}^{n} f(r)$. Under
this approximation, the singular points of $\mathcal{P}(H,r)$ in the
complex-$r$ plane are determined by the two expressions
\begin{equation}
    {\left( r_{c} - r \right)}^{n} = - \frac{H}{C_{1}}, \quad
    {\left( r_{c} - r \right)}^{n-1} = - \frac{H}{C_{2}}.
    \label{eq_blow}
\end{equation}
The full structure of the poles in Eq.~(\ref{eq_blow}) is detailed in the
Supplemental Material~\cite{SI}/\ref{appendix_poles}. However, as the integral
in Eq.~(\ref{eq_probhess_partint}) is performed over the real interval $[0,
r_c]$, the new upper limit is determined by the largest positive real root of
these expressions. This occurs at a value $r^{*} = r_{c} - s$, where $s \equiv
s(H)$ represents the shift in the upper limit.  The singular behaviour in
$\mathcal{P}(H,r)$ is thus determined by both the sign of $H$, and the signs of
the first two derivatives of the potential ($C_1$ and $C_2$), as the cut-off
distance $r \to r_{c}$ is approached.

\begin{table}[t!]
    \centering
    \caption{Asymptotic behaviour of Hessian element distributions in the limit
    $H \to 0$. The diagonal element distribution, depends on the relative signs
    of $H$ and the interaction potential near the cut-off $\psi_{\delta} \equiv
    \psi(r_{c}-\delta)$. The case $(*)$ corresponds to $H \times \psi_{\delta}
    > 0$, while $(\dagger)$ corresponds to $H \times \psi_{\delta} < 0$. The
    results are identical for both two and three dimensions.}\label{tab_results}
    \setlength{\tabcolsep}{8pt}
    \begin{tabular}[t]{lcl}
        \toprule
        Element $(H)$   & Smoothness $(n)$          & $\lim\limits_{H \to 0} P(H)$ \\
        \midrule
        Diagonal        & ${[2,\infty)}^{*}$        &
                        ${\left\lvert H \right\rvert}^{-1 + \frac{3}{2n}}$ \\
        \cmidrule(l){3-3}
        $(\alpha = \beta)$ & $ {\{2\}}^{\dagger} $       &
                        ${\left\lvert H \right\rvert}^{-1 + \frac{3}{2n}}$ \\
                        & ${\{3\}}^{\dagger}$       &
                        ${\left\lvert H \right\rvert}^{-\frac{1}{2}}$
                        $\log{\left({\left\lvert H \right\rvert}^{-1}\right)} $\\
                        & ${(3,\infty)}^{\dagger}$  &
                        ${\left\lvert H \right\rvert}^{-1 + \frac{1}{n-1}} $ \\
        \cmidrule(l){2-3}
        Off-diagonal    & $\{2\}$                   &
                        $\log{\left({\left\lvert H \right\rvert}^{-1}\right)} $ \\
        $(\alpha \neq \beta)$ & $(2,\infty)$              &
                        ${\left\lvert H \right\rvert}^{-1 + \frac{1}{n-1}}$ \\
        \bottomrule
    \end{tabular}
\end{table}

In order to analyse the asymptotic forms of $P(H)$ in
Eq.~(\ref{eq_probhess_partint}) as $H \to 0$, we define a small variable
$\epsilon$ as a distance to $r^{*}$ at a given value of $H \text{, } \epsilon =
(r_{c} - s) - r$. The derivatives of the potential described in
Eq.~(\ref{eq_approx}) can then be written as
\begin{equation}
    \psi_{r} / r \approx C_{1} {\left( \epsilon + s \right)}^{n}, \quad
    \psi_{r r} \approx C_{2} {\left( \epsilon + s \right)}^{n-1}
    \label{eq_limiting}.
\end{equation}
We can now extract the asymptotic behaviour of $P(H)$ using this approximation
in the partial integrands in
Eqs.~(\ref{eq_partint_2D})~and~(\ref{eq_partint_3D}). The behaviour of these
integrands depends on the relative signs of $H$ and the interaction potential
near cut-off $\psi_{\delta} \equiv \psi(r_{c}-\delta)$ with $\delta/r_c \ll 1$.
Below, we present the analysis for the diagonal elements in 2D, with the
details of all cases presented in the Supplemental
Material~\cite{SI}/\ref{appendix_asymptotes}. Using Eq.~(\ref{eq_limiting}) in
Eq.~(\ref{eq_partint_2D}), for $H \to 0, \; \epsilon \to 0$, we have
\begin{small}
\begin{equation}
    \mathcal{P}^{2\text{D}}_{\alpha \alpha} \left( H, \epsilon \right) \sim
    {\left\{ \left[ H + C_{1} {(\epsilon + s)}^{n} \right] \left[ H + C_{2}
    {(\epsilon + s)}^{n-1} \right] \right\}}^{-1/2}.
    \label{eq_asymp_Diag_2D}
\end{equation}
\end{small}
We can extract the limiting behaviour of $P(H)$ by identifying the dominant
contribution from the above expression to the integral in
Eq.~(\ref{eq_probhess_partint}).  There exist up to three regimes of $\epsilon$
in terms of the behaviour of the partial integrand in Eq.
(\ref{eq_asymp_Diag_2D}): (i) $\left[0, H^{1/n-1}\right)$, (ii)
$\left[H^{1/n-1}, H^{1/n}\right)$, and (iii) $\left[H^{1/n}, \infty\right)$,
depending on the shift $s(H)$ (refer to Supplemental
Material~\cite{SI}/\ref{appendix_poles}). For the case $H \times \psi_{\delta}
> 0$, contribution of the integral over the last interval dominates, with $P
\left( H \right) \sim \int_{H^{1/n}}^{\infty} \epsilon^{-n + 1/2} =
\left.{\epsilon^{-n + 3/2}}\right\vert_{H^{1/n}}^{\infty}$. Therefore, we
arrive at the asymptotic form $ P^{2\text{D}}_{\alpha \alpha} \left( H \right) \sim
H^{-1 + \frac{3}{2 n}}$. Our results for all cases are summarized in
Table~\ref{tab_results}. Remarkably, although the expressions in
Eqs.~(\ref{eq_partint_2D})~and~(\ref{eq_partint_3D}) have very different forms,
they yield exactly the same results for the singularities in both 2D and
3D.

\begin{figure}[t!]
    \resizebox{0.95\linewidth}{!}{\input{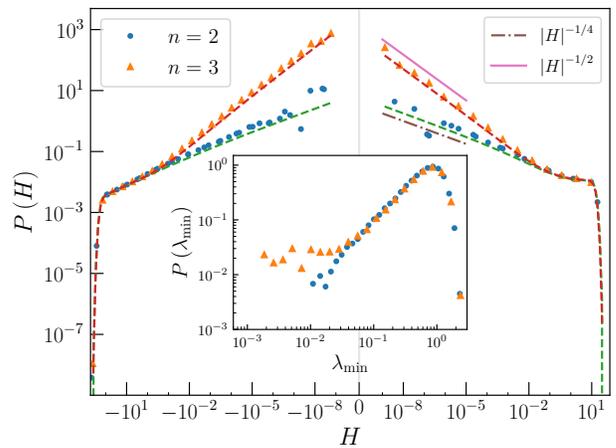}}
    \caption{Distribution of diagonal Hessian elements in the R10 system with
    different smoothness at cut-off $n=2$ and $3$ in two dimensions. The
    configurations sampled are at energy minima. The plot is in
    symmetric-log-log scale, showing the power law divergence in both positive
    and negative $H$. The dashed lines represent the analytic predictions from
    Eq.~(\ref{eq_probhess_partint}), using numerically sampled radial
    distribution functions. Exponent of the singularity changes with the
    degree of smoothness $n$.  (Inset) Distribution of the minimum eigenvalue
    $\lambda_{\min}$ of the Hessian (for system size $N=256$), displaying
    significant changes with the degree of smoothness
    $n$.}\label{fig_hess_distro}
\end{figure}

\textit{Numerical Simulations:}
In order to verify our predictions, we have performed extensive numerical
simulations of structural glass formers, in 2D as well as 3D. We simulate a
binary mixture of purely repulsive particles of type A and B (refer to
Supplemental Material~\cite{SI}/\ref{appendix_glass_models}).  The A-B
interactions are illustrated in Fig.~\ref{fig_countour}. High temperature
molecular dynamics simulations were utilised to generate independent,
uncorrelated configurations of particles, and sample their inherent structures
by locating the nearest local minimum via the conjugate-gradient minimization.
We then evaluate the Hessian elements between particles within interacting
range of each other. In Fig.~\ref{fig_hess_distro} we plot the numerically
sampled distributions of the diagonal Hessian elements for A-B interactions in
2D systems for $n =2$ and $n =3$, along with our theoretical predictions
(dashed lines) using Eq.~(\ref{eq_probhess_partint}) in
Eq.~(\ref{eq_partint_2D}), displaying near-perfect agreement. These
distributions diverge as $H \to 0^{+}$ with the exponents $-\frac{1}{4}$ for $n
=2$ and $-\frac{1}{2}$ for $n = 3$, as predicted in Table~\ref{tab_results}.

Finally, we turn our attention to the vibrational properties of the system,
that are probed through the eigenvalue spectrum of the Hessian matrix. We test
the sensitivity of low-lying eigenvalues to the smoothness of the potential,
and consequently, the power of the singularity in $P(H)$, by analysing the
distribution of the minimum eigenvalue, $\lambda_{\min}$. Our results for the
short-ranged R10 glass model, displayed in the inset of
Fig.~\ref{fig_hess_distro}, show a significant divergence at low values,
between distributions for two values of smoothness.

\textit{Discussion:}
We have presented analytic results for the distribution of Hessian elements in
disordered amorphous media in 2D and 3D, and verified them with extensive
numerical simulations. Our treatment is quite general, relying only on the
isotropy of the underlying amorphous medium, and can be extended to other
systems displaying such disorder. Additionally, we have shown that the Hessian
matrices of amorphous materials display a preponderance of small elements,
characterized by a singularity whose strength depends on the smoothness of the
interaction potential at the cut-off distance.  Remarkably, the results for the
singularities are exactly the same in both 2D and 3D, a fact that warrants
deeper investigation. We have also shown numerically that such singularities
have crucial implications for the low-lying eigenvalues of the Hessian matrix
that govern the stability or fragility of amorphous solids, highlighting their
sensitivity to the small, non-zero elements in the Hessian matrix. Our results
are particularly relevant for numerical studies of glasses, where the degree of
smoothness in interaction potentials have been shown to affect vibrational
properties~\cite{shimada2018anomalous}.  The limit where such interaction
potentials display sharp cut-offs are typically used in the study of jamming
transitions~\cite{o2003jamming,silbert2006structural, ramola2016disordered,
ramola2017scaling}, and it would be interesting to study the effect of
smoothness in the potential on the properties of such systems.  Finally, it
would also be interesting to extend our analytic results to construct bounds on
the vibrational density of states of amorphous systems~\cite{horn2012matrix}.

\textit{Acknowledgements:}
We thank Edan Lerner, Prathyush Manchala, Mustansir Barma, Daan Frenkel,
Srikanth Sastry, Chandan Dasgupta, Satya Majumdar, Kedar Damle and Deepak Dhar
for useful discussions.  V. V. K. thanks CSIR for the Shyama Prasad Mukherjee
Fellowship (SPM-07/1142(0228)/2015-EMR-1).  S. K. would like to acknowledge the
support from Swarna Jayanti Fellowship Grants No. DST/SJF/PSA-01/2018-19 and
No. SB/SFJ/2019-20/05.  This project was funded by intramural funds at TIFR
Hyderabad from the Department of Atomic Energy (DAE).

\nocite{plimpton1995fast, lammps}
\nocite{laug}
\nocite{intelmkl}
\nocite{mathematica}
\nocite{scipy}
\nocite{hunter2007matplotlib, matplotlib}

\bibliographystyle{apsrev4-2} 
\bibliography{Cusp_Hessian_Bibliography}

\clearpage

\begin{widetext}

\begin{appendix}

    \section*{\large \texorpdfstring{S\MakeLowercase{upplemental}
    M\MakeLowercase{aterial for}\\ ``C\MakeLowercase{usp singularities in}
    H\MakeLowercase{essian element distributions of amorphous media}''}
    {Supplemental Material for "Singularities in Hessian element
    distributions of amorphous media"}}

    In this document we provide supplemental figures and derivations related to
    the results presented in the main text. In Section~1, we provide detailed
    derivations of the partial integrands $\mathcal{P}(H,r)$ that are used in
    the computation of the Hessian element distributions $P(H)$ in two and
    three dimensions. In Section~2, we display a precise match between the
    analytic and numerical distributions obtained using an exponential pair
    correlation function $g_2(r)$. In Section~3, we provide details of the
    structure of the singularities of the partial integrand. In Section~4, we
    derive exact asymptotic forms for $P(H)$ in the limit $H \to 0$. Finally in
    Section~5, we provide details of the models and software used in our
    numerical simulations.

    \subsection{Hessian Element Distribution}~\label{appendix_exact_part_int}

    In this Section we derive exact forms for the partial integrands in both
    two and three dimensions as announced in
    Eqs.~(\ref{eq_partint_2D})~and~(\ref{eq_partint_3D}) in the main text. The
    elements of the Hessian matrix may be described in terms of the
    inter-particle distance vector ($\mathbf{r}^{i j}$), its magnitude ($r^{i
    j}$), and the pair-wise interaction potential ($\psi^{i j}$), which for
    central potentials is a function purely of the inter-particle distance
    $r^{i j}$ (Eq.~(\ref{eq_hess}) in main text)
    \begin{equation}
        \mathcal{H}_{\alpha \beta}^{i j} (\mathbf{r}^{i j}) = - \left(
        \frac{\psi_{r r}^{i j}}{{(r^{i j})}^{2}} -
        \frac{\psi_{r}^{i j}}{{(r^{i j})}^{3}} \right) r^{i j}_{\alpha} r^{i
        j}_{\beta} - \delta_{\alpha \beta} \frac{\psi_{r}^{i j}}{r^{i j}},
        \label{eq_sup_hess}
    \end{equation}
    where the subscripts of $r$ indicate partial derivatives $\psi_{r} =
    \partial_{r^{i j}} \psi$ and $\psi_{r r} = \partial_{r^{i j}}
    \partial_{r^{i j}} \psi$. Henceforth, we use $H$ and $r$ to represent the
    Hessian element and the inter-particle distance respectively. The
    distribution of the elements may then be computed as
    (Eq.~(\ref{eq_probhess}) in main text)
    \begin{equation}
        P \left( H \right) = \int d r \; d \Omega \; P \left( r, \Omega \right)
        \delta \left( H - \mathcal{H}_{\alpha \beta}^{i j} (\mathbf{r}^{i j})\right).
        \label{eq_sup_probhess}
    \end{equation}
    In addition, assuming a product form for the joint distribution of the
    inter-particle distances and angles $P(r,\Omega) = P(r)P(\Omega)$, we have
    (Eq.~(\ref{eq_probhess_partint}) in main text)
    \begin{equation}
        P \left( H \right) = \int_{0}^{r_{c}} d r \; P(r)  \; \frac{P \left(
        \Omega \right)}{\left\lvert \frac{\partial H}{\partial \Omega} \right\rvert}
        = \int_{0}^{r_{c}}d r \; P(r) \; \mathcal{P} (H, r).
        \label{eq_probhessDiag2D}
    \end{equation}
    In the sections that follow, we focus our attention on determining the
    partial integrand $\mathcal{P}(H,r)$.

    \subsubsection*{Diagonal elements}

    We first consider the diagonal elements $(\alpha = \beta)$ in two
    dimensions. Since by symmetry, the $xx$ and $yy$ elements are equivalent,
    we consider just the former.  The inter-particle distance vector is
    determined by its magnitude and angle ($\phi$) with respect to the
    $x$-axis. Using this we arrive at
    \begin{equation}
        \mathcal{H}_{x x}^{i j} \left(r,\cos \phi \right) = - \left(
        \psi_{r r}^{i j} - \frac{\psi_{r}^{i j}}{r} \right) \cos^{2} \phi
        -  \frac{\psi_{r}^{i j}}{r}.
        \label{eq_hessDiag2D}
    \end{equation}

    \begin{figure}[t!]
        \resizebox{0.45\textwidth}{!}{\input{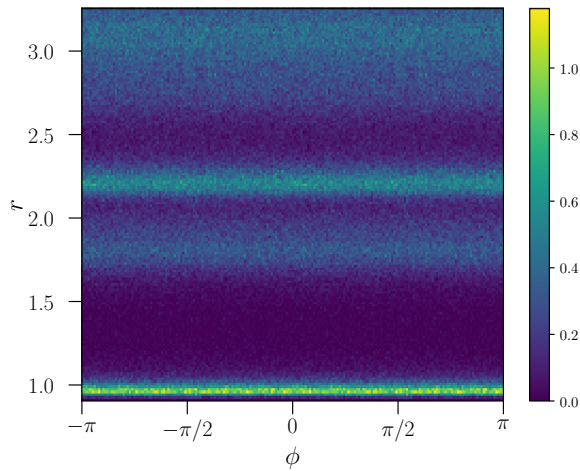}}
        \caption{Joint distribution of inter-particle distances ($r$) and
        angles ($\phi$) for energy minimised structures of an R10 system in two
        dimensions. This distribution is isotropic in the angular variable
        $\phi$, justifying a product form for this joint distribution.}
        \label{fig_rTheta_2dR10} 
    \end{figure}

    As stated above, we may infer from Fig.~\ref{fig_rTheta_2dR10} that the
    distributions of $r$ and $\phi$ are independent, and additionally, that the
    distribution in $\phi$ is uniform, with
    \begin{align}
        P(r,\phi) &= P(r) P(\phi), \nonumber\\
        P_{2\text{D}}(\phi) &= \frac{1}{2\pi}\label{eq_uncorr}.
    \end{align}
    Since the only dependence on $\phi$ is through a cosine, it is convenient
    to use the distribution
    \begin{equation}
        P_{2\text{D}}(\cos \phi) = \frac{1}{\pi \sqrt{1 - \cos \phi^{2}}}.
        \label{eq_probcos}
    \end{equation}
    The partial integrand in this case is
    \begin{equation}
        \mathcal{P}^{2\text{D}}_{\alpha \alpha} = {\left[ \left\lvert
        \frac{\partial H}{\partial \cos \phi} \right\rvert \sqrt{\left\lvert 1
        - \cos^{2} \phi \right\rvert} \right]}^{-1}.
        \label{eq_partint2D}
    \end{equation}
    We may use Eq.~(\ref{eq_hessDiag2D}) to simplify this, also noting that
    \begin{equation}
        \cos^{2} \phi = - \frac{H + \frac{\psi_{r}}{r}}{\psi_{r r} -
        \frac{\psi_{r}}{r}}.
        \label{eq_cosineDiag}
    \end{equation}
    Therefore, we have
    \begin{equation}
        \mathcal{P}^{2\text{D}}_{\alpha \alpha} \left( H, r \right) = {\left[ 2
        \sqrt{\left\lvert \left( H + \frac{\psi_{r}}{r} \right) \left( H +
        \psi_{r r} \right) \right\rvert} \right]}^{-1},
        \label{eq_sup_partint_Diag_2D}
    \end{equation}
    which is the first part of Eq.~(\ref{eq_partint_2D}) in the main text.

    We next derive the partial integrand in three dimensions. The
    inter-particle distance vector is now determined by its magnitude, the
    polar angle ($\theta$) subtended on the $z$-axis, and the azimuthal angle
    ($\phi$) subtended by the projection of the vector onto the $x-y$ plane,
    measured relative to the $x$-axis. By symmetry, the distributions of the
    $xx$, $yy$ and $zz$ Hessian elements are the same, and therefore we only
    need consider one of them
    \begin{equation}
        \mathcal{H}_{z z}^{i j} \left(r^{i j},\theta \right) = - \left(
        \psi_{r r}^{i j} - \frac{\psi_{r}^{i j}}{r} \right) \cos^{2} \theta
        -  \frac{\psi_{r}^{i j}}{r}.
        \label{eq_hessDiag3D}
    \end{equation}
    The partial integrand $\mathcal{P}(H,r)$ is then similar to that in two
    dimensions (Eq.~(\ref{eq_partint2D})), since the $zz$ element does not
    depend on the azimuthal angle $\phi$.  Next, we may infer from
    Fig.~\ref{fig_angular_3dR10} that the distribution of $\cos \theta$ is
    uniform
    \begin{equation}
        P_{3\text{D}}(\cos \theta) = \frac{1}{2}.
        \label{eq_probcos3D}
    \end{equation}
    Since the $zz$ element does not involve $\phi$, the integral over $\phi$ in
    Eq.~(\ref{eq_sup_probhess}) evaluates to a constant, and therefore, we
    arrive at the partial integrand
    \begin{equation}
        \mathcal{P}^{3\text{D}}_{\alpha \alpha} = {\left\lvert \frac{\partial
        H}{\partial \cos \theta} \right\rvert}^{-1}.
        \label{eq_above1}
    \end{equation}
    Solving Eq.~(\ref{eq_hessDiag3D}) for $\cos \theta$ as in
    Eq.~(\ref{eq_cosineDiag}), and substituting it in Eq. (\ref{eq_above1})
    above, we have
    \begin{equation}
        \mathcal{P}^{3\text{D}}_{\alpha \alpha} \left( H, r \right) = {\left[ 2
        \sqrt{\left\lvert \left( H + \frac{\psi_{r}}{r} \right) \left(\psi_{r
        r} - \frac{\psi_{r}}{r} \right) \right\rvert} \right]}^{-1},
        \label{eq_sup_partint_Diag_3D}
    \end{equation}
    which is the first part of Eq.~(\ref{eq_partint_3D}) in the main text.

    \begin{figure}[t!]
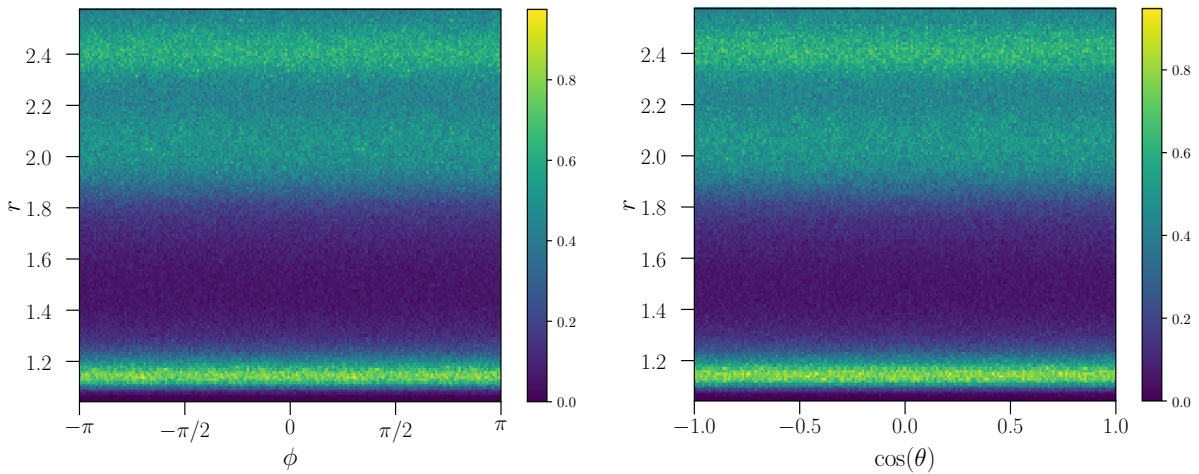

        \resizebox{0.45\textwidth}{!}{\input{rphi_3dR10.pgf}}
        \resizebox{0.45\textwidth}{!}{\input{rctheta_3dR10.pgf}}
        \caption{(Left) Joint distribution of inter-particle distances ($r$)
        and azimuthal angles ($\phi$) and (Right) joint distribution of
        inter-particle distances ($r$) and polar angles ($\cos \theta$) for
        energy minimised structures of an R10 system in three dimensions. In
        both cases the angular distributions are isotropic.}
        \label{fig_angular_3dR10} 
    \end{figure}

    \subsubsection*{Off-Diagonal elements}

    The off-diagonal elements  ($\alpha \neq \beta$)  of the Hessian in two
    dimensions may be represented as
    \begin{equation}
        \mathcal{H}_{x y}^{i j} \left(r, \phi \right) = - \left(
        \psi_{r r}^{i j} - \frac{\psi_{r}^{i j}}{r} \right) \cos \phi \sin \phi.
        \label{eq_hessOffDiag2D}
    \end{equation}
     Once again, the partial integrand takes the form
    \begin{equation}
        \mathcal{P}^{2\text{D}}_{\alpha \beta} = {\left[ \left\lvert
        \frac{\partial H}{\partial \cos \phi} \right\rvert \sqrt{\left\lvert 1
        - \cos^{2} \phi \right\rvert} \right]}^{-1},
        \label{eq_partint2D_2}
    \end{equation}
   which can be simplified to yield
    \begin{equation}
        \mathcal{P}^{2\text{D}}_{\alpha \beta} = {\left[ \left\lvert
        \frac{\partial H}{\partial \cos \phi} \sin \phi \right\rvert
        \right]}^{-1}.
        \label{eq_partintOffDiag2D1}
    \end{equation}
    Using Eq.~(\ref{eq_hessOffDiag2D}), the Jacobian factor is given by
    \begin{equation}
        \frac{\partial H}{\partial \cos \phi} = -\left( \psi_{r r} -
        \frac{\psi_{r}}{r} \right) \sin \phi + \left( \psi_{r r} -
        \frac{\psi_{r}}{r} \right) \frac{\cos^{2} \phi}{\sin \phi} = \left(
        \psi_{r r} - \frac{\psi_{r}}{r} \right) \frac{\cos 2\phi}{\sin \phi}.
    \end{equation}
    Substituting this back into Eq.~(\ref{eq_partintOffDiag2D1}), we arrive at
    \begin{equation}
        \mathcal{P}^{2\text{D}}_{\alpha \beta} = {\left[ \left\lvert \left(
        \psi_{r r} - \frac{\psi_{r}}{r} \right) \cos 2\phi \right\rvert
        \right]}^{-1}.
        \label{eq_partintOffDiag2D2}
    \end{equation}
    We may also rewrite Eq.~(\ref{eq_hessOffDiag2D}) as
    \begin{equation}
        \sin 2\phi = \frac{- 2 H}{\left( \psi_{r r} - \frac{\psi_{r}}{r}
        \right)}.
    \end{equation}
    Therefore
    \begin{equation}
        \cos 2\phi = \sqrt{1 - \frac{4 H^{2}}{{\left( \psi_{r r} -
        \frac{\psi_{r}}{r} \right)}^{2}}}.
    \end{equation}
    Thus, the partial integrand in terms of $H$ and $r$ simplifies to
    \begin{equation}
        \mathcal{P}^{2\text{D}}_{\alpha \beta} \left( H, r \right) = {\left[
            \sqrt{\left\lvert {\left( \psi_{r r} - \frac{\psi_{r}}{r}
            \right)}^{2} - 4 H^{2} \right\rvert} \right]}^{-1},
        \label{eq_sup_partint_OffDiag_2D}
    \end{equation}
    which is the second part of Eq.~(\ref{eq_partint_2D}) in the main text.

    Next, for the calculation for off-diagonal elements in three dimensions, it
    is convenient to choose $\mathcal{H}_{xz}$, which now additionally involves
    the polar angle $\theta$ as
    \begin{equation}
        \mathcal{H}_{x z}^{i j} \left(r, \theta, \phi \right) = - \left(
        \psi_{r r}^{i j} - \frac{\psi_{r}^{i j}}{r} \right) \cos \theta \sin \theta
        \cos \phi.
        \label{eq_hessOffDiag3D}
    \end{equation}
    The distribution of $\phi$ in 3D is similar to 2D, as can be seen from
    Fig.~\ref{fig_angular_3dR10}, with $P_{3\text{D}}(\phi) = \frac{1}{2 \pi}$.
    Therefore the partial integrand is given by
    \begin{equation}
        \mathcal{P}^{3\text{D}}_{\alpha \beta} = \int_{-1}^{1} d (\cos \theta)
        {\left[ \left\lvert \frac{\partial H}{\partial \cos \phi} \right\rvert
        \sqrt{\left\lvert 1 - \cos^{2} \phi \right\rvert} \right]}^{-1}.
        \label{eq_partintOffDiag3D1}
    \end{equation}
    The Jacobian in the above equation can be evaluated using
    Eq.~(\ref{eq_hessOffDiag3D}). Finally, substituting for $\cos \phi$ using
    Eq.~(\ref{eq_hessOffDiag3D}), we have
    \begin{equation}
        \mathcal{P}^{3\text{D}}_{\alpha \beta} = \int_{-1}^{1} d (\cos \theta)
        {\left[ \sqrt{ \left\lvert {\left( \psi_{r r} - \frac{\psi_{r}}{r}
        \right)}^{2} \sin^{2} \theta \cos^{2} \theta - H^{2} \right\rvert }
        \right]}^{-1}.
        \label{eq_partintOffDiag3D2}
    \end{equation}
    We may simplify this integral, and express it as
    \begin{equation}
        I \left( H ; \kappa \right) = \frac{\kappa}{H} \int^{1}_{-1}
        \frac{dx}{\sqrt{ x^{2} (1 - x^{2}) - \kappa^{2} }},
        \label{eq_wierd_integral}
    \end{equation}
    which is the second part of Eq.~(\ref{eq_partint_3D}) in the main text, and where
    \begin{equation}
        \kappa  = \frac{H}{\left( \psi_{r r} - \frac{\psi_{r}}{r} \right)}.
        \label{eq_above2}
    \end{equation}
    Since we are interested in the $H \to 0$ limit, we may assume $\kappa <
    1/2$. This guarantees that the term within the square root in Eq.
    (\ref{eq_wierd_integral}) remains real. The limits such that the integral
    in Eq.  (\ref{eq_wierd_integral}) is real, are given by
    \begin{equation}
        a = \sqrt{\frac{1-\sqrt{1-4 \kappa ^2}}{2}} \text{ and }
        b = \sqrt{\frac{1+\sqrt{1-4 \kappa ^2}}{2}}.
    \end{equation}
    We may then write
    \begin{equation}
        I \left( H ; \kappa \right) = \frac{\kappa}{H} \int^{b}_{a} \frac{2
        dx}{\sqrt{ x^{2} (1 - x^{2}) - \kappa^{2} }},
        \label{eq_wierd_integral2}
    \end{equation}
    Unfortunately, the integral above does not have a closed form solution,
    however we can still extract the asymptotic behaviour exactly. In order to
    make progress, let us re-write the integral described in
    Eq.~(\ref{eq_wierd_integral2}), by making the transformation
    \begin{equation}
        \kappa^{2} = \mu^{2} - \mu^{4},
        \label{eq_awesome_trick}
    \end{equation}
    which gives us
    \begin{equation}
        I \left( H ; \mu \right) = \frac{\sqrt{\mu^{2} - \mu^{4}}}{H}
        \int^{\sqrt{1 - \mu^{2}}}_{\mu} \frac{2 dx}{\sqrt{ x^{2} (1 - x^{2}) -
        \mu^{2} + \mu^{4} }},
        \label{eq_partintOffDiag3D3}
    \end{equation}
    where the limits simplify since we may write $x^2 \left( 1 - x^{2} \right)
    - \mu^{2} + \mu^{4} = \left( x^{2} - \mu^{2} \right) \left( \left(1 - x^{2}
    \right) - \mu^{2} \right)$.

    Next, we split the integrand into partial fractions as
    \begin{equation}
        \frac{1}{\sqrt{x^{2} (1 - x^{2}) - \mu^{2} + \mu^{4}}} =
        \frac{1}{\sqrt{x^{2} - \mu^{2}}} + \frac{1}{\sqrt{1 - x^{2} - \mu^{2}}}
        + \frac{1 - \sqrt{1 - x^{2} - \mu^{2}} - \sqrt{x^{2} - \mu^{2}}}{
            \sqrt{(1 - x^{2} - \mu^{2}) (x^{2} - \mu^{2})}}.
    \end{equation}
    We can now examine the integral over each of these terms separately. For
    the first term, we have
    \begin{align}
        \int^{\sqrt{1 - \mu^{2}}}_{\mu} \frac{dx}{\sqrt{ x^{2} - \mu^{2} }}
        &\hspace*{4 pt}= \left. \tanh^{-1}{\left(\frac{x}{\sqrt{x^{2} -
        \mu^{2}}}\right)} \right\lvert^{\sqrt{1 - \mu^{2}}}_{\mu} \nonumber \\
        &\hspace*{4 pt}= \tanh^{-1}{\left(\sqrt{\frac{\mu^{2} - 1}{2 \mu^{2} -
        1}}\right)} + i \frac{\pi}{2} \nonumber \\
        &\overset{\mu \to 0}{\approx} \log{(2)} - \log{(\mu )} +
        O\left(\mu^{2}\right).
        \label{eq_partintOffDiag3D3_term1}
    \end{align}
    Note that this is singular at $\mu = 0$. Next, the integral over the second
    term yields
    \begin{align}
        \int^{\sqrt{1 - \mu^{2}}}_{\mu} \frac{dx}{\sqrt{ 1- x^{2} - \mu^{2} }}
        &\hspace*{4 pt}= \left. \tan^{-1}{\left(\frac{x}{\sqrt{1 - x^{2} -
        \mu^{2}}}\right)} \right\lvert^{\sqrt{1 - \mu^{2}}}_{\mu} \nonumber \\
        &\hspace*{4 pt}=  \frac{\pi}{2} - \tan^{-1}{\left(\frac{\mu}{\sqrt{1 -
        2 \mu^{2}}}\right)} \nonumber \\
        &\overset{\mu \to 0}{\approx} \frac{\pi}{2} - \mu +
        O\left(\mu^{3}\right).
        \label{eq_partintOffDiag3D3_term2}
    \end{align}
    The integral over the third term cannot be performed exactly, so we first
    examine its behaviour as $\mu \to 0$
    \begin{equation}
        \frac{1 - \sqrt{1 - x^{2} - \mu^{2}} - \sqrt{x^{2} - \mu^{2}}}{
            \sqrt{(1 - x^{2} - \mu^{2}) (x^{2} - \mu^{2})}} \overset{\mu \to
        0}{\approx} - \frac{1 + x - \sqrt{1 - x^{2}}}{x (1 + x)} +
        O\left(\mu^{2}\right).
    \end{equation}
    We therefore have
    \begin{align}
        \int^{\sqrt{1 - \mu^{2}}}_{\mu} \frac{(-1 - x + \sqrt{1 - x^{2}})dx}{x
        (1 + x)}
        &\hspace*{4 pt}= \sin^{-1}{(\mu)} - \cos^{-1}{(\mu )} - \log
        \left(\frac{\mu + 1}{\sqrt{1 - \mu^{2}} + 1}\right) \nonumber \\
        &\overset{\mu \to 0}{=} \log{(2)} - \frac{\pi}{2}.
        \label{eq_partintOffDiag3D3_term3}
    \end{align}
    Putting together the terms from
    Eqs.~(\ref{eq_partintOffDiag3D3_term1}),~(\ref{eq_partintOffDiag3D3_term2})
    and (\ref{eq_partintOffDiag3D3_term3}), into
    Eq.~(\ref{eq_partintOffDiag3D3}) and ignoring terms of $O(\mu)$, we have
    \begin{equation}
        I \left( H ; \mu \right) \approx  \frac{2 \sqrt{\mu^{2} - \mu^{4}}}{H}
        \left[ \log{(4)} - \log{(\mu)} + O\left(\mu\right) \right].
    \end{equation}
    Thus the singular behaviour of the full integrand occurs due to the first
    term in the partial fractions. Finally, we transform back to $\kappa$ using
    the appropriate solution, i.e., the positive real branch in
    Eq.~(\ref{eq_awesome_trick}) given by
    \begin{equation}
        \mu = \frac{\sqrt{1-\sqrt{1-4 \kappa^{2}}}}{\sqrt{2}}
    \end{equation}
    Therefore $I(H;\kappa)$ (Eq.~(\ref{eq_wierd_integral})), in the limit of
    small $\kappa$, simplifies to
    \begin{equation}
        I \left( H ; \kappa \right) \approx  \frac{2 \kappa}{H}
        \left[ \log{(4)} - \log{(\kappa)}  + O\left(\kappa^{2}\right)\right]
    \end{equation}
    Thus, the partial integrand for the off-diagonal elements in three
    dimensions, when $\kappa \ll 1$ is given by
    \begin{equation}
        \mathcal{P}^{3\text{D}}_{\alpha \beta} \left( H, r \right) = 2 {\left(
        \psi_{r r} - \frac{\psi_{r}}{r} \right)}^{-1} \log{\left( \frac{4
        \left(\psi_{r r} - \frac{\psi_{r}}{r} \right)}{H} \right)},
        \label{eq_sup_partint_OffDiag_3D}
    \end{equation}
    which is Eq.~(\ref{eq_partint_limit_OffDiag_3D}) in the main text.


    \begin{figure}[t!]
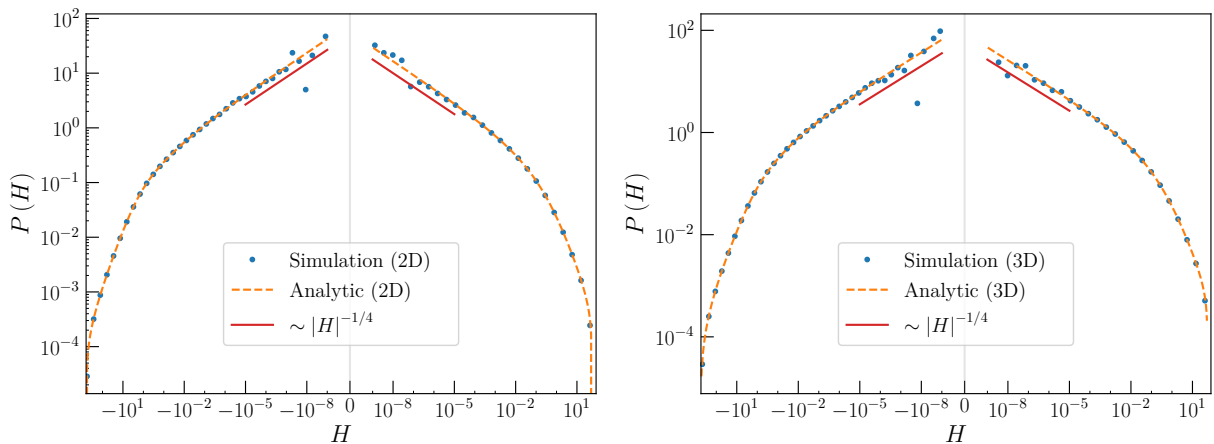

        \resizebox{0.45\textwidth}{!}{\input{hessian_distributions_ExpG2RShift.pgf}}
        \resizebox{0.45\textwidth}{!}{\input{hessian_distributions_ExpG2RShift3D.pgf}}
        \caption{Distribution of diagonal Hessian elements ($\alpha = \beta$)
        corresponding to A-B interactions in the R10 model, using an exact form
        for the radial distribution function $g_2(r)$ given in
        Eq.~(\ref{eq_exact_g2}) for (Left) two dimensional systems and (Right)
        three dimensional systems.}
        \label{fig_hess_distro_expg2rshift} 
    \end{figure}

    \subsection{\texorpdfstring{$P(H)$}{P(H)} using an exact radial distribution function} 
    \label{appendix_exact_g2r} 

    The analytical prediction of the Hessian element distribution as shown in
    Fig.~\ref{fig_hess_distro} in the main text contains slight discrepancies
    which occur due to the interpolated form of the $g_2(r)$ used. In this
    Section we analyse the distribution of Hessian elements sampled from an
    artificial, but exact distribution of pair distances given by
    \begin{equation}
        g_{2}(r) = e^{-(r-1)} \Theta(r-1),
    \label{eq_exact_g2}
    \end{equation}
    where $\Theta(r)$ is the Heaviside theta function. Our results for diagonal
    Hessian elements ($\alpha = \beta$) are displayed in
    Fig.~\ref{fig_hess_distro_expg2rshift} for both two and three dimensional
    systems. We find that the analytic procedure yields a prediction that
    agrees with the numerical distribution to within numerical precision.

    \subsection{Poles in the complex-\texorpdfstring{$r$}{r} plane}\label{appendix_poles}

    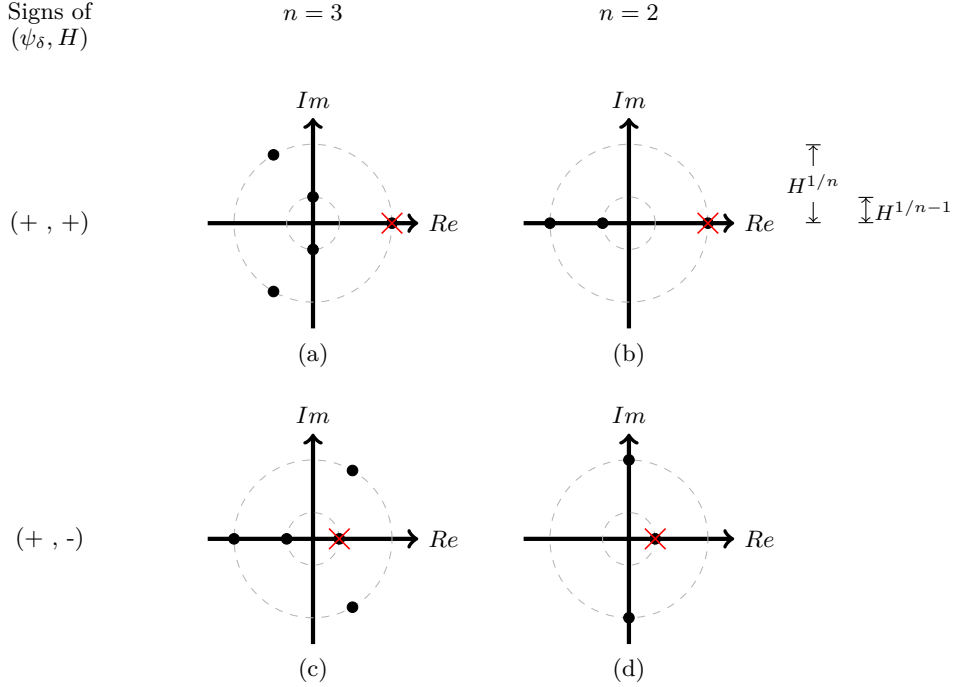
\begin{figure}[t!]
        \centering
        \usetikzlibrary{shapes.misc}
        \tikzset{cross/.style={cross out, draw=red, minimum
        size= 2 * (#1-\pgflinewidth), inner sep=0pt, outer sep=0pt},
        cross/.default={1}}
        \begin{tikzpicture}[scale=0.7]
        \pgfmathsetmacro{\RADIUSB}{1.5}
        \pgfmathsetmacro{\RADIUSS}{0.5}
        \pgfmathtruncatemacro{\SMOOTH}{3}
        \pgfmathtruncatemacro{\SMOOTHA}{\SMOOTH - 1}
        \pgfmathtruncatemacro{\SMOOTHB}{\SMOOTH - 2}
        \pgfmathtruncatemacro{\SMOOTHC}{\SMOOTH - 3}
        \draw node at (-8,-5) {Signs of};
        \draw node at (-8,-5.5) {($\psi_{\delta} , H$)};
        \draw node at (-3,-5) {$n = \SMOOTH$};
        \draw node at (3,-5) {$n = \SMOOTHA$};
        \draw[|<->|] (6.5,-9)--(6.5,-9+\RADIUSB)
            node[pos=0.5,font=\footnotesize,fill=white] {$H^{1/n}$};
        \draw[|<->|] (7.5,-9)--(7.5,-9+\RADIUSS)
            node[pos=0.5,anchor=west,font=\footnotesize] {$H^{1/n-1}$};
        \draw node at (-8,-9) {(+ , +)};
        \draw[->,ultra thick] (-5,-9)--(-1,-9) node[right]{$Re$};
        \draw[->,ultra thick] (-3,-11)--(-3,-7) node[above]{$Im$};
        \draw[help lines, color=gray!60, dashed] (-3,-9) circle (\RADIUSS);
        \draw[help lines, color=gray!60, dashed] (-3,-9) circle (\RADIUSB);
        \foreach \x in {0,...,\SMOOTHA} 
            \filldraw ({\RADIUSB*cos(2*pi*\x/\SMOOTH r)-3},
            {\RADIUSB*sin(2*pi*\x/\SMOOTH r)-9}) circle (0.1);
        \foreach \x in {0,...,\SMOOTHB} 
            \filldraw ({\RADIUSS*cos(((2*\x)+1)*pi/\SMOOTHA r)-3},
            {\RADIUSS*sin(((2*\x)+1)*pi/\SMOOTHA r)-9}) circle (0.1);
        \draw[semithick] (-3+\RADIUSB,-9) node[cross=4.5] {};
        \draw node at (-3,-11.5) {(a)};
        \draw[->,ultra thick] (1,-9)--(5,-9) node[right]{$Re$};
        \draw[->,ultra thick] (3,-11)--(3,-7) node[above]{$Im$};
        \draw[help lines, color=gray!60, dashed] (3,-9) circle (\RADIUSS);
        \draw[help lines, color=gray!60, dashed] (3,-9) circle (\RADIUSB);
        \foreach \x in {0,...,\SMOOTHB} 
            \filldraw ({\RADIUSB*cos(2*pi*\x/\SMOOTHA r)+3},
            {\RADIUSB*sin(2*pi*\x/\SMOOTHA r)-9}) circle (0.1);
        \foreach \x in {0,...,\SMOOTHC} 
            \filldraw ({\RADIUSS*cos(((2*\x)+1)*pi/\SMOOTHB r)+3},
            {\RADIUSS*sin(((2*\x)+1)*pi/\SMOOTHB r)-9}) circle (0.1);
        \draw[semithick] (3+\RADIUSB,-9) node[cross=4.5] {};
        \draw node at (3,-11.5) {(b)};
        \draw node at (-8,-15) {(+ , -)};
        \draw[->,ultra thick] (-5,-15)--(-1,-15) node[right]{$Re$};
        \draw[->,ultra thick] (-3,-17)--(-3,-13) node[above]{$Im$};
        \draw[help lines, color=gray!60, dashed] (-3,-15) circle (\RADIUSS);
        \draw[help lines, color=gray!60, dashed] (-3,-15) circle (\RADIUSB);
        \foreach \x in {0,...,\SMOOTHA} 
            \filldraw ({\RADIUSB*cos(((2*\x)+1)*pi/\SMOOTH r)-3},
            {\RADIUSB*sin(((2*\x)+1)*pi/\SMOOTH r)-15}) circle (0.1);
        \foreach \x in {0,...,\SMOOTHB} 
            \filldraw ({\RADIUSS*cos(2*pi*\x/\SMOOTHA r)-3},
            {\RADIUSS*sin(2*pi*\x/\SMOOTHA r)-15}) circle (0.1);
        \draw[semithick] (-3+\RADIUSS,-15) node[cross=4.5] {};
        \draw node at (-3,-17.5) {(c)};
        \draw[->,ultra thick] (1,-15)--(5,-15) node[right]{$Re$};
        \draw[->,ultra thick] (3,-17)--(3,-13) node[above]{$Im$};
        \draw[help lines, color=gray!60, dashed] (3,-15) circle (\RADIUSS);
        \draw[help lines, color=gray!60, dashed] (3,-15) circle (\RADIUSB);
        \foreach \x in {0,...,\SMOOTHB} 
            \filldraw ({\RADIUSB*cos(((2*\x)+1)*pi/\SMOOTHA r)+3},
            {\RADIUSB*sin(((2*\x)+1)*pi/\SMOOTHA r)-15}) circle (0.1);
        \foreach \x in {0,...,\SMOOTHC} 
            \filldraw ({\RADIUSS*cos(2*pi*\x/\SMOOTHB r)+3},
            {\RADIUSS*sin(2*pi*\x/\SMOOTHB r)-15})
            circle (0.1);
        \draw[semithick] (3+\RADIUSS,-15) node[cross=4.5] {};
        \draw node at (3,-17.5) {(d)};
        \end{tikzpicture}
        \caption{An illustration of the structure of poles of
        Eq.~(\ref{eq_sup_blow}) in the complex-$(r_c - r)$ plane.  The figure
        shows the poles for degrees of smoothness $n = 2$ and $n = 3$. The
        behaviour depends on the relative signs of the potential near the
        cut-off $\left( \psi_{\delta} \equiv \psi(r_{c}-\delta) \right)$ and
        the Hessian element $H$.  The (red) crosses indicate the largest real
        positive pole that controls the upper limit of the integral in
        Eq.~(\ref{eq_sup_partint_Diag_2D}). The shift $s(H)$ is given by the
        distance of this pole to the origin.}
        \label{fig_poles} 
    \end{figure}

    In this Section we analyse the variation of the upper limit in $r$ that
    appears in the integral expression for $P(H)$ in
    Eq.~(\ref{eq_probhessDiag2D}) (Eq.~(\ref{eq_probhess_partint}) in the main
    text). For example, in the case $H > 0$, the contours plotted in the inset
    of Fig.~\ref{fig_countour} in the main text indicate a maximum value of $r$
    (i.e., the turn-around points of the contours). This implies a divergence
    stemming from the Jacobian factor in the integral, which can be extracted
    from the roots of the denominator of $\mathcal{P}(H,r)$ given in
    Eqs.~(\ref{eq_partint_2D})~and~(\ref{eq_partint_3D}).  Below, we will focus
    on the non-trivial case of diagonal elements in 2D
    (Eq.~(\ref{eq_sup_partint_Diag_2D})). The other cases are treated in a
    similar manner, and are significantly simpler.

    As stated in Eq.~(\ref{eq_approx}) in the main text, for a given degree of
    smoothness $n$ at the cut-off $r_{c}$, the interaction potential and its
    derivatives may be approximated in terms  of the approach to $r_{c}$ as
    \begin{align}
        \psi \left(r\right) &= {\left( r_{c} - r \right)}^{n+1} f \left(r\right)
        \nonumber \\
        \psi_{r} / r &\approx C_{1} {\left( r_{c} - r \right)}^{n} \nonumber \\
        \psi_{r r} &\approx C_{2} {\left( r_{c} - r \right)}^{n-1},
        \label{eq_sup_approx}
    \end{align}
    where $f(r)$ is a regular function, along with $C_{1} = - (n+1) f(r) / r$,
    and $C_{2} = n (n+1) f(r)$ which vary significantly slower compared to the
    power-law term as $r \to r_{c}$. The singular behaviour in the partial
    integrand in Eq.~(\ref{eq_sup_partint_Diag_2D}) occurs under one of the
    following conditions:
    \begin{align}
        {\left( r_{c} - r \right)}^{n} &= - \frac{H}{C_{1}} \nonumber \\
        {\left( r_{c} - r \right)}^{n-1} &= - \frac{H}{C_{2}}.
        \label{eq_sup_blow}
    \end{align}
    Using this approximation, for given a value of $H$, the value of $r$ at
    which the partial integrand is singular is \emph{shifted} from the cut-off,
    which we quantify as
    \begin{equation}
        r^{*}(H) = r_{c} - s(H).
        \label{eq_singular_shift}
    \end{equation}
    The shifts $s(H)$ depend on the relative signs of the potential near
    cut-off $\left( \psi_{\delta} \equiv \psi(r_{c}-\delta) \text{ with }
    \delta/r_c \ll 1 \right)$, and the value of the Hessian element $H$. For
    the sake of convenience, let us denote the signs of these two quantities as
    a pair: $\left( \psi_{\delta}, H \right)$.

    The value of $r$ at which the partial integrand $\mathcal{P}(H,r)$ becomes
    singular is determined by the largest, real, positive pole in
    Eq.~(\ref{eq_sup_blow}). The shift $s(H)$ can now be identified based on
    the fact that positive unity always has a positive root, while negative
    unity does not. The structure of the poles of the partial integrand in 2D
    (for $n = 2$ and $3$) are displayed in Fig.~\ref{fig_poles}, with the
    positive real roots that contribute to the shift denoted by (red) crosses.
    The shifts for these cases have been summarized in  Table~\ref{tab_shifts}.

    \begin{table}[ht]
        \centering
        \setlength{\tabcolsep}{8pt}
        \begin{tabular}[t]{cl}
            \toprule
            Signs of    & Shift \\
            $\left( \psi_{\delta}, H \right)$ & $s(H)$ \\
            \midrule
            $(+,+)$     & ${(H/C_{1})}^{1/n}$ \\
            $(+,-)$     & ${(H/C_{2})}^{1/n-1}$ \\
            $(-,+)$     & ${(H/C_{2})}^{1/n-1}$ \\
            $(-,-)$     & ${(H/C_{1})}^{1/n}$ \\
            \bottomrule
        \end{tabular}
        \caption{The various possible shifts $s(H)$ in the upper limit of the
        integral in Eq.~(\ref{eq_probhessDiag2D}), for the diagonal elements
        $(\alpha = \beta)$ in two dimensions. The value of the shift depends on
        the relative signs of $H \text{ and } \psi_{\delta}$ as well as the
        degree of smoothness $n$.}\label{tab_shifts}
    \end{table}


    \subsection{Asymptotic Integrals}\label{appendix_asymptotes}

    The behaviour of the partial integrand $\mathcal{P}(H,r)$ as one approaches
    the upper limit $r^*(H)$, controls the asymptotic form of $P(H)$.  It is
    therefore convenient to parametrise these expressions in terms of the
    distance $\epsilon$ to this singular value $r^*(H)$, with
    \begin{equation}
        \epsilon = r^{*}(H) - r = r_{c} - s(H) - r.
        \label{eq_approach}
    \end{equation}
    The partial integrands display different regimes of decay in $\epsilon$
    which are controlled by the distance to the singularities in the
    complex-$r$ plane as described in the previous section.  For example, in
    the case of $n = 2$, Fig~\ref{fig_part_int} displays the variation of the
    partial integrand in Eq.~(\ref{eq_sup_partint_Diag_2D}) at three different
    values of $H = 10^{-2}, 10^{-3}$ and $10^{-4}$.  In this case, there are
    two regimes of decay, with the point of turn-around scaling as
    $\sqrt{\lvert H \rvert}$, as may be seen in the scaling collapse displayed
    in the inset.

    \begin{figure}[t]
        \resizebox{0.5\textwidth}{!}{\input{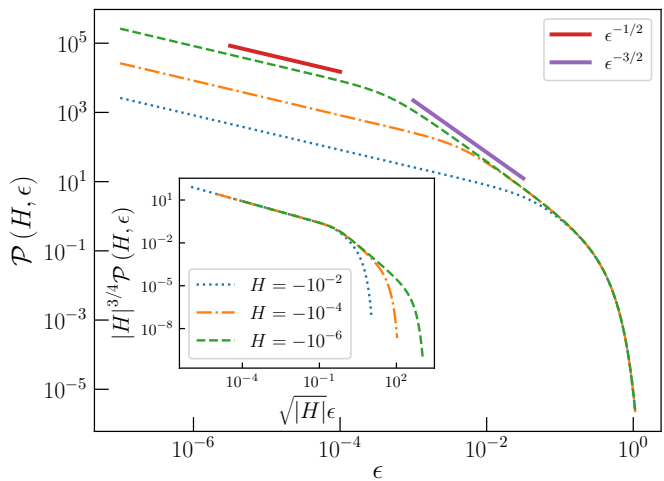}}
        \caption{Plot of the asymptotic partial integrand
        $\mathcal{P}(H,\epsilon)$ given in
        Eq.~(\ref{eq_sup_negH_asymp_Diag_2D}), corresponding to A-B
        interactions of the R10 model with degree of smoothness $n =2$. The two
        regimes of asymptotic behaviour are given by $[0,\sqrt{H})$ and
        $[\sqrt{H},\infty)$, with decays $\epsilon^{-1/2}$ and
        $\epsilon^{-3/2}$ respectively, as predicted in
        Eqs.~(\ref{eq_sup_n2_integral2})~and~(\ref{eq_sup_diag_n2_integral1}).
        (Inset) Scaling collapse of the integrand with the scaled variable
        $\sqrt{\lvert H \rvert} \epsilon$.}
        \label{fig_part_int} 
    \end{figure}

    We now examine each of the partial integrands derived in
    Sec.~\ref{appendix_exact_part_int}, in the limit of the inter-particle
    distances approaching the singular value $r^*(H)$, i.e. $\epsilon \ll 1$,
    as described in Sec.~\ref{appendix_poles}.  We first derive limiting forms
    for the partial integrands, identifying the various regimes of decay in
    $\epsilon$. Finally we use these asymptotic forms to perform the
    integration in Eq.~(\ref{eq_probhessDiag2D}), to extract the dominant
    behaviour of $P(H)$ as $H \to 0$. In terms of the distance $\epsilon$ and
    the shift $s \equiv s(H)$, we may rewrite Eq.~(\ref{eq_sup_approx}) as
    \begin{equation}
        \psi_{r} / r \approx C_{1} {\left( \epsilon + s \right)}^{n}, \quad
        \psi_{r r} \approx C_{2} {\left( \epsilon + s \right)}^{n-1},
        \label{eq_sup_limiting}
    \end{equation}
    which is Eq.~(\ref{eq_limiting}) in the main text.  In the rest of this
    section, we focus on the case when the potential is positive near cut-off
    $(\psi_{\delta} > 0)$, as is the case for the R10 model that we study
    numerically (displayed in Fig.~\ref{fig_countour} in the main text). The
    case for $(\psi_{\delta} < 0)$ proceeds exactly in the same manner, with $H
    \to -H$. Note that the signs of the derivatives of the potential near the
    cut-off, alternate with the order of derivative. This may be seen from
    Eq.~(\ref{eq_sup_approx}), and implies that $\psi_{\delta}, C_{1} \text{
        and } C_{2}$ have alternately opposing signs. This allows us to
    determine the shifts from Eq.~(\ref{eq_sup_blow}), purely in terms of the
    signs of $H$, given the sign of $\psi_{\delta}$.

    \subsubsection*{Diagonal elements}

    \textbf{Two Dimensions:~}
    For the case of diagonal elements in two dimensions, inserting
    Eq.~(\ref{eq_sup_limiting}) in Eq.~(\ref{eq_sup_partint_Diag_2D}), we
    have
    \begin{equation}
        \mathcal{P}^{2\text{D}}_{\alpha \alpha} \left( H, \epsilon \right) \sim
        {\left\{ \left[ H + C_{1} {(\epsilon + s)}^{n} \right] \left[ H + C_{2}
        {(\epsilon + s)}^{n-1} \right] \right\}}^{-1/2}.
        \label{eq_sup_asymp_Diag_2D}
    \end{equation}
    As detailed in Section~\ref{appendix_poles}, the value of the shift $s(H)$
    depends on the sign of $H$, i.e., the direction of approach to $0$.

    \textbf{(a)}
    Let us first consider the case of $H < 0$. As shown in
    Fig.~\ref{fig_poles}~(c,d) and Table~\ref{tab_shifts}, for this case, $s
    \sim {\lvert H \rvert}^{1/(n-1)}$.  Eq.~(\ref{eq_sup_asymp_Diag_2D}),
    ignoring an overall scale factor, may then be written as
    \begin{equation}
        \mathcal{P}^{2\text{D}}_{\alpha \alpha} \left( H < 0, \epsilon \right)
        \sim {\left\{ \left[ - \lvert H \rvert - {\left(\epsilon + {\lvert H
        \rvert}^{\frac{1}{n-1}}\right)}^{n} \right] \left[ - \lvert H \rvert +
        {\left(\epsilon + {\lvert H \rvert}^{\frac{1}{n-1}}\right)}^{n-1}
        \right] \right\}}^{-\frac{1}{2}}.
        \label{eq_sup_negH_asymp_Diag_2D}
    \end{equation}
    At small values of $\epsilon$, this asymptotic form simplifies into
    power-laws in $\epsilon$. The number of power-law regimes present are
    governed by the distance to the poles in the complex-$\epsilon$ plane. As
    an illustrative example, let us focus on the case of $n = 2$. The structure
    of the poles for this case have been illustrated in
    Fig.~\ref{fig_poles}~(d). Here, all the poles are at the same distance
    scale $\sqrt{\lvert H \rvert}$ away from $\epsilon = 0$, implying two
    regimes of decay in $\epsilon$, separated by this scale. We illustrate this
    in Fig.~\ref{fig_part_int} where we plot the asymptotic partial integrand
    in Eq.~(\ref{eq_sup_negH_asymp_Diag_2D}). The inset shows a scaling
    collapse of the partial integrand for different values of $H$, displaying
    the $\sqrt{\lvert H  \rvert}$ scale in the turning point.

    Finally, substituting Eq.~(\ref{eq_sup_negH_asymp_Diag_2D}) into the
    integral for $P(H)$ (Eq.~(\ref{eq_probhessDiag2D})), we can identify the
    contributions of each of the slopes to the asymptotic behaviour of $P(H)$ as
    $H \to 0$. Below, we present these simplifications and integrations for each
    regime, with the dominant terms marked with boxes.  For the case of $n=2$,
    we split the integral in Eq.~(\ref{eq_sup_negH_asymp_Diag_2D}) into two
    parts $I_1(H)$ and $I_2(H)$, arising from the regions $\epsilon \in
    [\sqrt{H},\infty)$ and $\epsilon \in [0,\sqrt{H})$ respectively.  In the
    first case, the singular contribution to $P(H)$ is given by
    \begin{equation}
        I_{1} \left( H \right) \sim \int_{\sqrt{H}}^{\infty}
        {\left[ (\epsilon^{2}) (\epsilon) \right]}^{-\frac{1}{2}}
        = \int_{\sqrt{H}}^{\infty} \epsilon^{-\frac{3}{2}}
        = \left.\epsilon^{-\frac{1}{2}}\right\rvert^{\infty}_{\sqrt{H}}
        \sim \boxed{H^{-\frac{1}{4}}}.
        \label{eq_sup_diag_n2_integral1}
    \end{equation}
    Similarly, when $\epsilon \in [0,\sqrt{H})$, the singular contribution is
    given by
    \begin{equation}
        I_{2} \left( H \right) \sim \int_{0}^{\sqrt{H}}
        {\left[ (H) (\epsilon) \right]}^{-\frac{1}{2}}
        = H^{-\frac{1}{2}} \int_{0}^{\sqrt{H}} \epsilon^{-\frac{1}{2}}
        = H^{-\frac{1}{2}} \left.\epsilon^{\frac{1}{2}}\right\rvert_{0}^{\sqrt{H}}
        = H^{-\frac{1}{4}}.
        \label{eq_sup_n2_integral2}
    \end{equation}
    Note that the analysis above, predicts power laws of $\epsilon^{-3/2}$ and
    $\epsilon^{-1/2}$ of the integrand, as displayed in
    Fig.~\ref{fig_part_int}. Thus, both regimes contribute to the dominant
    singular behaviour in $P(H)$ which has the singularity $H^{-\frac{1}{4}}$,
    as displayed in Fig.~\ref{fig_prob_2D_diag}. Next, for $n \ge 3$, an
    additional (smaller) distance scale also appears, namely $(\sim {\lvert H
    \rvert}^{1/n-1})$, as shown  in Fig.~\ref{fig_poles}~(c).  This implies
    three regimes of decay in $\epsilon$, separated by the two scales ${\lvert
    H \rvert}^{1/n-1}$ and ${\lvert H \rvert}^{1/n}$.  Once again we can
    evaluate the contribution from each of these regimes to the singular
    behaviour in $P(H)$ and identify the dominant singularity.

    We treat the case $n =3$ separately, as this presents logarithmic
    singularities as we show below. In this case we split the integral in
    Eq.~(\ref{eq_sup_negH_asymp_Diag_2D}) into three parts
    \begin{equation}
        I_{1} \left( H \right) \sim \int_{{H}^{\frac{1}{3}}}^{\infty}
        {\left[ (\epsilon^{3}) (\epsilon^{2}) \right]}^{-\frac{1}{2}}
        = \int_{{H}^{\frac{1}{3}}}^{\infty} \epsilon^{-\frac{5}{2}}
        = \left.\epsilon^{-\frac{3}{2}}\right\rvert_{{H}^{\frac{1}{3}}}^{\infty}
        = H^{-\frac{1}{2}}.
    \end{equation}
    \begin{equation}
        I_{2} \left( H \right) \sim \int_{{H}^{\frac{1}{2}}}^{{H}^{\frac{1}{3}}}
        {\left[ (H) (\epsilon^{2}) \right]}^{-\frac{1}{2}}
        = H^{-\frac{1}{2}} \int_{{H}^{\frac{1}{2}}}^{{H}^{\frac{1}{3}}} \epsilon^{-1}
        = H^{-\frac{1}{2}} \left. {\log{\epsilon}}
        \right\rvert_{{H}^{\frac{1}{2}}}^{{H}^{\frac{1}{3}}}
        \sim \boxed{H^{-\frac{1}{2}} \log{H^{-1}}}.
        \label{eq_sup_n3_integral2}
    \end{equation}
    \begin{equation}
        I_{3} \left( H \right) \sim \int_{0}^{\sqrt{H}}
        {\left[ (H) (\epsilon H^{\frac{1}{2}}) \right]}^{-\frac{1}{2}}
        = H^{-\frac{3}{4}} \int_{0}^{\sqrt{H}} \epsilon^{-\frac{1}{2}}
        = H^{-\frac{3}{4}} \left.\epsilon^{\frac{1}{2}}\right\rvert_{0}^{\sqrt{H}}
        = H^{-\frac{1}{2}}.
    \end{equation}
    Therefore the dominant singular behaviour in $P(H)$ for $n=3$ is
    accompanied by a logarithmic singularity. This behaviour has been displayed
    using results from simulations in Fig.~\ref{fig_prob_2D_diag}.  Next, we
    derive results for any general $n > 3$. Once again, splitting the integral
    in Eq.~(\ref{eq_sup_negH_asymp_Diag_2D}) into three parts, we have
    \begin{equation}
        I_{1} \left( H \right) \sim \int_{{H}^{\frac{1}{n}}}^{\infty}
        {\left[ (\epsilon^{n}) (\epsilon^{n-1}) \right]}^{-\frac{1}{2}}
        = \int_{{H}^{\frac{1}{n}}}^{\infty} \epsilon^{-n + \frac{1}{2}}
        = \left.\epsilon^{-n + \frac{3}{2}}\right\rvert_{{H}^{\frac{1}{n}}}^{\infty}
        = H^{-1 + \frac{3}{2n}}.
    \end{equation}
    \begin{equation}
        I_{2} \left( H \right) \sim \int_{{H}^{\frac{1}{n-1}}}^{{H}^{\frac{1}{n}}}
        {\left[ (H) (\epsilon^{n-1}) \right]}^{-\frac{1}{2}}
        = H^{-\frac{1}{2}} \int_{{H}^{\frac{1}{n-1}}}^{{H}^{\frac{1}{n}}}
        \epsilon^{-\frac{n}{2}+\frac{1}{2}}
        = H^{-\frac{1}{2}} \left. {\epsilon^{-\frac{n}{2} + \frac{3}{2}}}
        \right\rvert_{{H}^{\frac{1}{n-1}}}^{{H}^{\frac{1}{n}}}
        \sim \boxed{H^{-1 + \frac{1}{n-1}}}.
        \label{eq_sup_n_integral2}
    \end{equation}
    \begin{equation}
        I_{3} \left( H \right) \sim \int_{0}^{{H}^{\frac{1}{n-1}}}
        {\left[ (H) (\epsilon H^{\frac{n-2}{n-1}}) \right]}^{-\frac{1}{2}}
        = H^{-\frac{2n-3}{2n-2}} \int_{0}^{{H}^{\frac{1}{n-1}}} \epsilon^{-1/2}
        = H^{-\frac{2n-3}{2n-2}} \left. \epsilon^{1/2}
        \right\rvert_{0}^{{H}^{\frac{1}{n-1}}}
        \sim H^{-1 + \frac{1}{n-1}}.
    \end{equation}
    Above we have simplified the expressions in $I_{3}$ via a binomial
    expansion in the second term of Eq.~(\ref{eq_sup_negH_asymp_Diag_2D}). In
    this case, we find two competing powers arising from the three regions. The
    boxed result is dominant for $n>3$, whereas the contribution of the first
    region is dominant in the  $n \le 3$ cases as we have detailed above. We
    verify this non-trivial \emph{switching} behaviour with increasing degree
    of smoothness $n$ with results from simulations that are displayed in
    Fig.~\ref{fig_prob_2D_diag}.

    \begin{figure}[t!]
        \resizebox{0.5\textwidth}{!}{\input{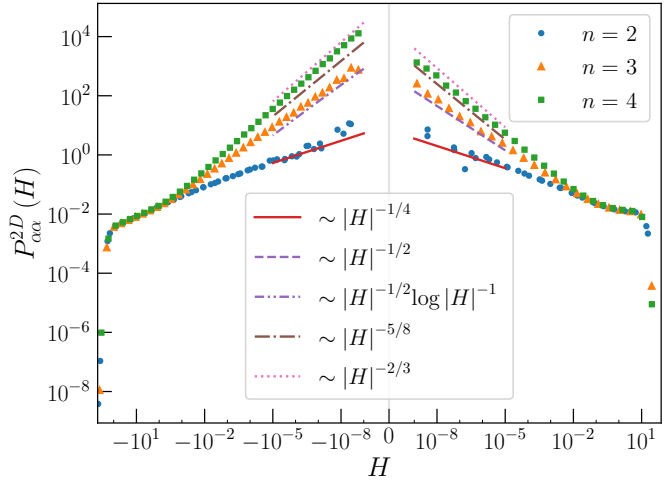}}
        \caption{Numerically sampled distribution (points) of diagonal Hessian
        elements $(\alpha = \beta)$ in energy minimized configurations of the
        R10 system in two dimensions, for various degrees of smoothness ($n$)
        of the potentials at cut-off. The plots show Hessian elements
        corresponding only to A-B interactions. The asymptotic behaviour of the
        singularities match our theoretical predictions (lines) given in
        Eqs.~(\ref{eq_sup_diag_n2_integral1}),~(\ref{eq_sup_n3_integral2})
        and~(\ref{eq_sup_n_integral2}).}
        \label{fig_prob_2D_diag} 
    \end{figure}

    \textbf{(b)}
    Next, we consider the case of $H > 0$. In this case, the pole structure
    displayed in Fig.~\ref{fig_poles}~(a,b) and detailed in
    Table~\ref{tab_shifts}, yields a shift $s \sim {\lvert H \rvert}^{1/n}$.
    Once again, Eq.~(\ref{eq_sup_asymp_Diag_2D}), up to an overall scale
    factor, can be simplified to yield
    \begin{equation}
        \mathcal{P}^{2\text{D}}_{\alpha \alpha} \left( H > 0, \epsilon \right)
        \sim {\left\{ \left[ \lvert H \rvert - {\left(\epsilon + {\lvert H
        \rvert}^{\frac{1}{n}}\right)}^{n} \right] \left[ \lvert H \rvert +
        {\left(\epsilon + {\lvert H \rvert}^{\frac{1}{n-1}}\right)}^{n-1}
        \right] \right\}}^{-\frac{1}{2}}.
        \label{eq_sup_posH_asymp_Diag_2D}
    \end{equation}
    In this case, there is only one distance scale away from the real positive
    pole: $\sim {\lvert H \rvert}^{1/n}$, and consequently only two regimes of
    decay in $\epsilon$. Following the same procedure as for the case of $H <
    0$, we split the integral in Eq.~(\ref{eq_probhessDiag2D}) into two parts,
    yielding the asymptotic behaviours
    \begin{equation}
        I_{1} \left( H \right) \sim \int_{{H}^{\frac{1}{n}}}^{\infty}
        {\left[ (\epsilon^{n}) (\epsilon^{n-1}) \right]}^{-\frac{1}{2}}
        = \int_{{H}^{\frac{1}{n}}}^{\infty} \epsilon^{-n + \frac{1}{2}}
        = \left.\epsilon^{-n + \frac{3}{2}}\right\rvert_{{H}^{\frac{1}{n}}}^{\infty}
        = \boxed{H^{-1 + \frac{3}{2n}}}.
    \end{equation}
    \begin{equation}
        I_{2} \left( H \right) \sim \int_{0}^{{H}^{\frac{1}{n}}}
        {\left[ (\epsilon H^{\frac{n-1}{n}})(H^{\frac{n-1}{n}}) \right]}^{-\frac{1}{2}}
        = H^{-1+\frac{1}{2n}} \int_{0}^{{H}^{\frac{1}{n}}} \epsilon^{-1/2}
        = H^{-1+\frac{1}{2n}} \left. \epsilon^{1/2}
        \right\rvert_{0}^{{H}^{\frac{1}{n}}}
        \sim H^{-1 + \frac{3}{2n}}.
    \end{equation}
    Above, the expression in $I_{2}$ has been simplified using a binomial
    expansion of the first term in Eq.~(\ref{eq_sup_posH_asymp_Diag_2D}). In
    this case we find that both regions contribute the same power $H^{-1 +
    \frac{3}{2n}}$, as corroborated with simulation results displayed in
    Fig.~\ref{fig_prob_2D_diag}.  A striking aspect of our analysis, is the
    prediction and numerical confirmation of the differences in powers of the
    singularities, on the negative and positive branches of $P(H)$ for the
    cases $n \ge 3$.  On the positive side $H > 0$ of $P(H)$, there are no
    changes in the slope scenarios, with the degree of smoothness $n$.

    \textbf{Three Dimensions:~}
    We next consider the distribution of diagonal Hessian elements in three
    dimensions. Inserting Eq.~(\ref{eq_sup_limiting}) into the expression for
    the partial integrand in Eq.~(\ref{eq_sup_partint_Diag_3D}), we arrive at
    \begin{equation}
        \mathcal{P}^{3\text{D}}_{\alpha \alpha} \left( H, \epsilon \right) \sim
        {\left\{ \left[ H + C_{1} {(\epsilon + s)}^{n} \right] \left[C_{2}
        {(\epsilon + s)}^{n-1} - C_{1} {(\epsilon + s)}^{n} \right]
        \right\}}^{-1/2}.
        \label{eq_sup_asymp_Diag_3D}
    \end{equation}

    \begin{figure}[t!]
        \resizebox{0.5\textwidth}{!}{\input{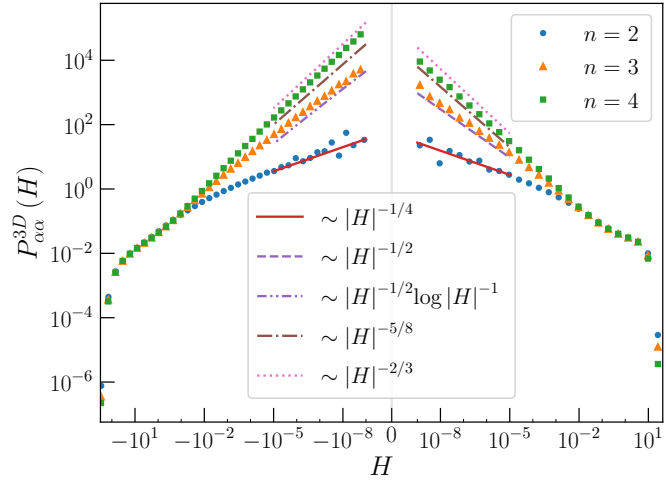}}
        \caption{Numerically sampled distribution (points) of diagonal Hessian
        elements $(\alpha = \beta)$ in energy minimized configurations of the
        R10 system in three dimensions, for various degrees of smoothness ($n$)
        of the potentials at cut-off. The plots show Hessian elements
        corresponding only to A-B interactions. The asymptotic behaviour of the
        singularities match our theoretical predictions (lines) obtained from
        Eqs.~(\ref{eq_sup_n2_integral1}),~(\ref{eq_sup_3d_diag_cases})
        and~(\ref{eq_sup_n_integral3}).}
        \label{fig_prob_3D_diag} 
    \end{figure}

    \textbf{(a)}
    As in two-dimensions, let us begin by considering the case of $H < 0$.
    However, unlike in two dimensions, we now have only one `ring' of poles,
    since only the first term in Eq.~(\ref{eq_sup_asymp_Diag_3D}) contains an
    $H$. In this case, there are no poles on the positive real axis, implying a
    shift $s = 0$. However, an `effective' shift in the limits of the integral
    in Eq.~(\ref{eq_probhessDiag2D}) still occurs in order to maintain a real
    valued integrand. The limits of the integral in $r$ are thus determined by
    imposing the condition $\cos^{2} \theta \leq 1$ in
    Eq.~(\ref{eq_hessDiag3D}). This may also be inferred from the $H < 0$
    contours in the inset of Fig.~\ref{fig_countour} in the main text.  Using
    the asymptotic approximations (Eq.~(\ref{eq_sup_approx})) in the afore
    mentioned condition yields a shift in the integration limit of $s(H) \sim
    {\lvert H \rvert}^{1/(n-1)}$.  We therefore arrive at
    \begin{equation}
        \mathcal{P}^{3\text{D}}_{\alpha \alpha} \left( H < 0, \epsilon \right)
        \sim {\left\{ \left[ - \lvert H \rvert - {\left(\epsilon + {\lvert H
        \rvert}^{\frac{1}{n-1}}\right)}^{n} \right] \left[ {\left(\epsilon +
        {\lvert H \rvert}^{\frac{1}{n-1}}\right)}^{n-1} + {\left(\epsilon +
        {\lvert H \rvert}^{\frac{1}{n-1}}\right)}^{n} \right]
        \right\}}^{-\frac{1}{2}}.
        \label{eq_sup_negH_asymp_Diag_3D}
    \end{equation}
    Next, let us examine the different regimes of behaviour of the partial
    integrand in $\epsilon$. The first region in this case is $\epsilon \in [
        {\lvert H \rvert}^{1/n},\infty )$, yielding a contribution
    \begin{equation}
        I_{1} \left( H \right) \sim \int_{{H}^{\frac{1}{n}}}^{\infty}
        {\left[ (\epsilon^{n}) (\epsilon^{n-1}) \right]}^{-\frac{1}{2}}
        = \int_{{H}^{\frac{1}{n}}}^{\infty} \epsilon^{-n + \frac{1}{2}}
        = \left.\epsilon^{-n + \frac{3}{2}}\right\rvert_{{H}^{\frac{1}{n}}}^{\infty}
        = \boxed{H^{-1 + \frac{3}{2n}}}.
        \label{eq_sup_n2_integral1}
    \end{equation}
    The second region is $\epsilon \in [ {\lvert H \rvert}^{1/(n-1)}, {\lvert H
    \rvert}^{1/n} )$, which yields a contribution (for $n \neq 3$)
    \begin{equation}
        I_{2} \left( H \right) \sim \int_{{H}^{\frac{1}{n-1}}}^{{H}^{\frac{1}{n}}}
        {\left[ (H) (\epsilon^{n-1}) \right]}^{-\frac{1}{2}}
        = H^{-\frac{1}{2}} \int_{{H}^{\frac{1}{n-1}}}^{{H}^{\frac{1}{n}}}
        \epsilon^{-\frac{n}{2}+\frac{1}{2}}
        = H^{-\frac{1}{2}} \left. {\epsilon^{-\frac{n}{2} + \frac{3}{2}}}
        \right\rvert_{{H}^{\frac{1}{n-1}}}^{{H}^{\frac{1}{n}}}
        = H^{-\frac{1}{2}} \left( H^{-\frac{n-3}{2n}} - H^{-\frac{n-3}{2(n-1)}} \right).
    \end{equation}
    Note that for the case $n =3$, the above integral yields a logarithmic
    singularity. The dominant term in the integral therefore depends on the
    value of the smoothness $n$ as
    \begin{align}
        I_{2} \left( H \right) \sim
        \begin{cases}
            {\left\lvert H \right\rvert}^{-1 + \frac{3}{2n}} &  n \in \text{$[2,3)$} \\
            \boxed{{\left\lvert H \right\rvert}^{\frac{1}{2}} \log{H^{-1}}} &  n = 3 \\
            {\left\lvert H \right\rvert}^{-1 + \frac{1}{n-1}} & n \in (3,\infty).
        \end{cases}
        \label{eq_sup_3d_diag_cases}
    \end{align}
    The third region is given by $\epsilon \in [ 0, {\lvert H \rvert}^{1/(n-1)}
    )$, with a contribution
    \begin{equation}
        I_{3} \left( H \right) \sim \int_{0}^{{H}^{\frac{1}{n-1}}}
        {\left[ (H) (H) \right]}^{-\frac{1}{2}}
        = H^{-1} \int_{0}^{{H}^{\frac{1}{n-1}}} d \epsilon
        = H^{-1} \left. \epsilon \right\rvert_{0}^{{H}^{\frac{1}{n-1}}}
        \sim \boxed{H^{-1 + \frac{1}{n-1}}}.
        \label{eq_sup_n_integral3}
    \end{equation}
    Once again, the behaviour of the dominant slope follows that of $I_{2}$
    (Eq.~(\ref{eq_sup_3d_diag_cases})). This behaviour is illustrated in the
    results from simulations shown in Fig.~\ref{fig_prob_3D_diag}.

    \textbf{(b)}
    We next turn to the positive branch of the distribution diagonal elements
    in 3D when $H >0$. Similar to the case of two dimensions, the shift is
    calculated from the first term in Eq.~(\ref{eq_sup_asymp_Diag_3D}) and is
    given by $s \sim {\lvert H \rvert}^{1/n}$, allowing the expression to be
    simplified to
    \begin{equation}
        \mathcal{P}^{3\text{D}}_{\alpha \alpha} \left( H > 0, \epsilon \right)
        \sim {\left\{ \left[ \lvert H \rvert - {\left(\epsilon + {\lvert H
        \rvert}^{\frac{1}{n-1}}\right)}^{n} \right] \left[ {\left(\epsilon +
        {\lvert H \rvert}^{\frac{1}{n-1}}\right)}^{n-1} + {\left(\epsilon +
        {\lvert H \rvert}^{\frac{1}{n-1}}\right)}^{n} \right]
        \right\}}^{-\frac{1}{2}}.
        \label{eq_sup_posH_asymp_Diag_3D}
    \end{equation}
    Similar to the corresponding case in two dimensions, there are no changes
    in the slopes with the smoothness. The distance scale is now given by $\sim
    {\lvert H \rvert}^{1/n}$. Splitting the integral in
    Eq.~(\ref{eq_probhessDiag2D}) into two regions, we have the contributions
    \begin{equation}
        I_{1} \left( H \right) \sim \int_{{H}^{\frac{1}{n}}}^{\infty}
        {\left[ (\epsilon^{n}) (\epsilon^{n-1}) \right]}^{-\frac{1}{2}}
        = \int_{{H}^{\frac{1}{n}}}^{\infty} \epsilon^{-n + \frac{1}{2}}
        = \left.\epsilon^{-n + \frac{3}{2}}\right\rvert_{{H}^{\frac{1}{n}}}^{\infty}
        = \boxed{H^{-1 + \frac{3}{2n}}}.
    \end{equation}
    \begin{equation}
        I_{2} \left( H \right) \sim \int_{0}^{{H}^{\frac{1}{n}}}
        {\left[ (\epsilon H^{\frac{n-1}{n}})(H^{\frac{n-1}{n}}) \right]}^{-\frac{1}{2}}
        = H^{-1+\frac{1}{2n}} \int_{0}^{{H}^{\frac{1}{n}}} \epsilon^{-1/2}
        = H^{-1+\frac{1}{2n}} \left. \epsilon^{1/2}
        \right\rvert_{0}^{{H}^{\frac{1}{n}}}
        \sim H^{-1 + \frac{3}{2n}}.
    \end{equation}
    The first term in $I_{2}$, has been simplified via a binomial expansion of
    the second term in Eq.~(\ref{eq_sup_posH_asymp_Diag_3D}).  In this case,
    both slopes contribute the same power $H^{-1 + \frac{3}{2n}}$,as shown in
    results from simulations in Fig.~\ref{fig_prob_3D_diag}. As in two
    dimensions, there is a difference in slopes on the negative and positive
    branches of $P(H)$ in three dimensions.

    \begin{figure}[t!]
        \resizebox{0.5\textwidth}{!}{\input{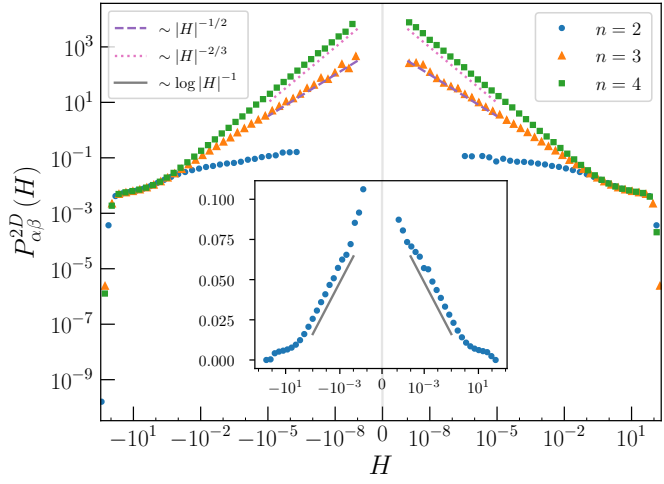}}
        \caption{Numerically sampled distribution (points) of off-diagonal
        Hessian elements $(\alpha \neq \beta)$ in energy minimized
        configurations of the R10 system in two dimensions, for various degrees
        of smoothness ($n$) of the potentials at cut-off. The plots show
        Hessian elements corresponding only to A-B interactions. The asymptotic
        behaviour of the singularities match our theoretical predictions
        (lines) obtained from Eq.~(\ref{eq_sup_2d_offdiag_cases}).  (Inset)
        Semi-log plot of the distribution for $n=2$, highlighting the
        logarithmic singularity.}
        \label{fig_prob_2D_offdiag} 
    \end{figure}

    \subsubsection*{Off-Diagonal elements}

    \textbf{Two Dimensions:~}
    Next, we turn to the distribution of off-diagonal Hessian elements in two
    dimensions, with the partial integrand given in
    Eq.~(\ref{eq_sup_partint_OffDiag_2D}). It is clear from this expression
    that $H < 0$ and $H > 0$, have exactly the same behaviour as the element
    $H$ only appears as a quadratic form.  Using Eq.~(\ref{eq_sup_limiting}) in
    the partial integrand for off-diagonal elements
    (Eq.~(\ref{eq_sup_partint_OffDiag_2D})), we have
    \begin{equation}
        \mathcal{P}^{2\text{D}}_{\alpha \beta} \left( H, \epsilon \right) \sim
        {\left\{ {\left[C_{2} {(\epsilon + s)}^{n-1} - C_{1} {(\epsilon +
        s)}^{n} \right]}^{2} - 4 H^{2} \right\}}^{-1/2}.
        \label{eq_sup_asymp_OffDiag_2D}
    \end{equation}
    Computing the relevant pole in the above expression, we find that the shift
    $s \sim {\lvert H \rvert}^{1/(n-1)}$. Therefore, the partial integrand, up
    to an overall scale factor is given by
    \begin{equation}
        \mathcal{P}^{2\text{D}}_{\alpha \beta} \left( H, \epsilon \right) \sim
        {\left\{ {\left[ {\left( \epsilon + {\lvert H \rvert}^{\frac{1}{n-1}}
        \right)}^{n-1} + {\left( \epsilon + {\lvert H \rvert}^{\frac{1}{n-1}}
        \right)}^{n} \right]}^{2} - H^{2} \right\}}^{-\frac{1}{2}}.
        \label{eq_sup_nonH_asymp_OffDiag_2D}
    \end{equation}
    Once again we compute the powers of the singularity in $P(H)$.  In this
    case the partial integrand has three regimes of decay, that can be
    extracted by appropriate binomial expansions of
    Eq.~(\ref{eq_sup_nonH_asymp_OffDiag_2D}), yielding the regions (i) $[
        {\lvert H \rvert}^{1/(n-1)}, \infty )$, (ii) $[ {\lvert H
        \rvert}^{2/(n-1)}, {\lvert H \rvert}^{1/(n-1)} )$ and (iii) $[ 0,
        {\lvert H \rvert}^{2/(n-1)} )$ In region (i), we have
    \begin{equation}
        I_{1} \left( H \right) \sim \int_{{H}^{\frac{1}{n-1}}}^{\infty}
        {\left[ {(\epsilon^{n-1})}^{2} \right]}^{-\frac{1}{2}}
        = \int_{{H}^{\frac{1}{n-1}}}^{\infty} \epsilon^{-n+1}
        = \left. \epsilon^{-n+2} \right\rvert_{{H}^{\frac{1}{n-1}}}^{\infty},
    \end{equation}
    yielding the following singular behaviours
    \begin{align}
        I_{1} \left( H \right) \sim
        \begin{cases}
            \boxed{\log{{\left\lvert H \right\rvert}^{-1}}} &  n = 2 \\
            \boxed{{\left\lvert H \right\rvert}^{-1 + \frac{1}{n-1}}}
            & n \in (2,\infty).
        \end{cases}
        \label{eq_sup_2d_offdiag_cases}
    \end{align}
    The integral in the second region $\epsilon \in [ {\lvert H
    \rvert}^{2/(n-1)}, {\lvert H \rvert}^{1/(n-1)} )$ can be performed by
    binomially expanding  the terms of $(\epsilon + s)$. Therefore the
    contribution from this region is
    \begin{equation}
        I_{2} \left( H \right) \sim \int_{{H}^{\frac{2}{n-1}}}^{{H}^{\frac{1}{n-1}}}
        {\left[ {\left( H + \epsilon H^{\frac{n-2}{n-1}} \right) + \left(
        H^{\frac{n}{n-1}} \right)}^{2} - H^{2} \right]}^{-\frac{1}{2}}
        \sim H^{\frac{-(2n-3)}{2(n-1)}} \int_{ {H}^{\frac{2}{n-1}} }^{
            {H}^{\frac{1}{n-1}} } \epsilon^{-\frac{1}{2}}
        \sim H^{-1 + \frac{1}{n-1}}
    \end{equation}
    The contribution from the third region $\epsilon \in [ 0, {\lvert H
    \rvert}^{2/(n-1)} )$ can be evaluated as
    \begin{equation}
        I_{3} \left( H \right) \sim \int_{0}^{{H}^{\frac{2}{n-1}}}
        {\left[ {\left( H + H^{\frac{n}{n-1}} \right)}^{2} - H^{2}
        \right]}^{-\frac{1}{2}}
        \sim H^{\frac{-(2n-1)}{2(n-1)}} \int_{0}^{
            {H}^{\frac{2}{n-1}} } d\epsilon
        \sim H^{-1 + \frac{3}{2(n-1)}}
    \end{equation}
    Therefore the dominant contribution arises from the first two regions
    $(I_{1}, I_{2})$.  Additionally our analysis predicts, a logarithmic
    singularity for the case $n=2$. These behaviours are illustrated in our
    numerically sampled distributions of off-diagonal elements in two
    dimensions in Fig.~\ref{fig_prob_2D_offdiag}.

    \begin{figure}[t!]
        \resizebox{0.5\textwidth}{!}{\input{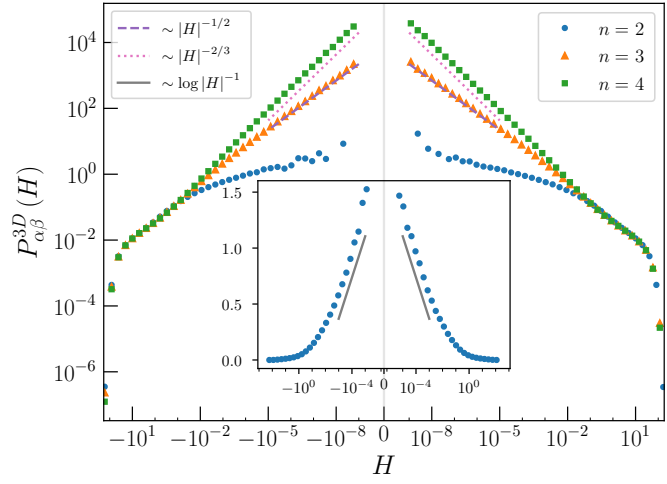}}
        \caption{Numerically sampled distribution (points) of off-diagonal
        Hessian elements $(\alpha \neq \beta)$ in energy minimized
        configurations of the R10 system in three dimensions, for various
        degrees of smoothness of the potentials at cut-off. The plots show
        Hessian elements corresponding only to A-B interactions. The asymptotic
        behaviour of the singularities match our theoretical predictions
        (lines) obtained from Eq.~(\ref{eq_sup_3d_offdiag_cases}). Semi-log
        plot of the distribution for $n=2$, highlighting the logarithmic
        singularity.}
        \label{fig_prob_3D_offdiag} 
    \end{figure}

    \textbf{Three Dimensions:~}
    Finally, to find the asymptotic behaviour of off-diagonal elements in three
    dimensions, we insert Eq.~(\ref{eq_sup_limiting}) into the expression for
    the partial integrand in the limit $\kappa = H {\left( \psi_{r r} -
    \frac{\psi_{r}}{r} \right)}^{-1} \to  0$ in
    Eq.~(\ref{eq_sup_partint_OffDiag_3D}). We therefore arrive at (ignoring an
    overall multiplicative constant)
    \begin{equation}
        \mathcal{P}^{3\text{D}}_{\alpha \beta} \left( H, \epsilon \right) \sim
        {\left( \epsilon^{n-1} + \epsilon^{n} \right)}^{-1} \log{ \left\lvert
        \frac{\epsilon^{n-1} - \epsilon^{n}}{H} \right\rvert }.
        \label{eq_sup_asymp_OffDiag_3D}
    \end{equation}
    In this case, the limits of the integral in Eq.~(\ref{eq_probhessDiag2D})
    are determined by the region of validity $(\kappa < 1)$ of the partial
    integrand Eq.~(\ref{eq_wierd_integral}). Since we are working with the
    asymptotic expression Eq.~(\ref{eq_sup_partint_OffDiag_3D}), there is only
    one region to consider, and we have the following contribution to the
    Hessian element distribution
    \begin{align}
        I \left( H \right) &\sim \int_{{H}^{\frac{1}{n-1}}}^{\infty}
        \epsilon^{-n+1} \log{\left(H^{-1} \epsilon^{n-1}\right)} \nonumber \\
        &\sim \left.{\frac{\epsilon^{-n+2}}{{(n-2)}^{2}} \left[ (n-1) - (n-2)
        \log{\left(H^{-1} \epsilon^{n-1}\right)} \right]}
        \right\vert^{\infty}_{H^{\frac{1}{n-1}}}
    \end{align}
    We therefore obtain the following asymptotic behaviour
    \begin{align}
        I \left( H \right) \sim
        \begin{cases}
            \boxed{\log{{\left\lvert H \right\rvert}^{-1}}} &  n = 2 \\
            \boxed{{\left\lvert H \right\rvert}^{-1 + \frac{1}{n-1}}}
            & n \in (2,\infty).
        \end{cases}
        \label{eq_sup_3d_offdiag_cases}
    \end{align}
    These behaviours have been illustrated in the results from our numerical
    simulations in Fig.~\ref{fig_prob_3D_offdiag}. The case of $n=2$ yields a
    logarithmic singularity, as may be seen from the inset of the figure.

    In summary, we have analysed the singularities associated with the
    distributions of Hessian elements for all cases in two and three
    dimensions, for any degree of smoothness $n$ of the interaction potential
    at the cut-off distance. Remarkably we find the same asymptotic behaviours
    in the distribution of Hessian elements for \emph{all} cases in \emph{both}
    two and three dimensions. This is surprising since the mechanisms leading
    to these singularities detailed above are very different. We note that
    although the dominant singularity exhibits this universal feature, the
    sub-dominant corrections to these behaviours may be different in two and
    three dimensions.  The power of the singularities of the distributions of
    diagonal elements display a change in behaviour at $n = 3$, and the
    off-diagonal elements, at $n=2$. In both cases, systems at the cross-over
    value of smoothness, display a logarithmic singularity in the distribution
    of Hessian elements.


    \subsection{Additional Simulation Details}
    \label{appendix_glass_models} 

    \subsubsection*{Simulation Potentials}

    We simulate a 50:50 mixture of two particle types A and B. The interaction
    potentials are cut-off at a distance
    \begin{equation}
        r_{c} = 1.385418025 \; \sigma.
    \end{equation}
    with the three interaction diameters given by
    \begin{align}
        \sigma_{A A} &= 1.0, \nonumber \\
        \sigma_{B B} &= 1.4, \nonumber \\
        \sigma_{A B} &= \sqrt{\sigma_{A A}\sigma_{B B}}.
    \end{align}
    The only difference between the parameters in two and three dimensions of
    this model are the reduced densities given by
    \begin{align}
        \rho_{2\text{D}} &= 0.85, \nonumber \\
        \rho_{3\text{D}} &= 0.81.
    \end{align}

    In our simulations we focus on the purely repulsive pairwise potential,
    given by a tenth order polynomial, termed `R10'. For the case of degree of
    smoothness $n = 2$, the potential is given by
    \begin{align}
        \psi &= {\left(\frac{\sigma}{r}\right)}^{10} + c_{0} + c_{2}
        {\left(\frac{r}{\sigma}\right)}^{2} + c_{4}
        {\left(\frac{r}{\sigma}\right)}^{4}, \nonumber \\
        &c_{0} = -0.8061409035399235, \nonumber \\
        &c_{2} = +0.7, \nonumber \\
        &c_{4} = -0.15630021928760743.
    \end{align}
    In the case of the potential smoothed to three derivatives $(n =3)$, we have
    \begin{align}
        \psi = {\left(\frac{\sigma}{r}\right)}^{10} &+ c_{0}
        + c_{2} {\left(\frac{r}{\sigma}\right)}^{2}
        + c_{4} {\left(\frac{r}{\sigma}\right)}^{4}
        + c_{6} {\left(\frac{r}{\sigma}\right)}^{6}, \nonumber \\
        c_{0} &= -2.149709076106463, \nonumber \\
        c_{2} &= +2.8000000000000007, \nonumber \\
        c_{4} &= -1.2504017543008594, \nonumber \\
        c_{6} &= +0.1900092331616435.
    \end{align}
    Finally, in the case of the potential smoothed to four derivatives $(n=4)$,
    we have
    \begin{align}
        \psi = {\left(\frac{\sigma}{r}\right)}^{10} + c_{0}
        &+ c_{2} {\left(\frac{r}{\sigma}\right)}^{2}
        + c_{4} {\left(\frac{r}{\sigma}\right)}^{4}
        + c_{6} {\left(\frac{r}{\sigma}\right)}^{6}
        + c_{8} {\left(\frac{r}{\sigma}\right)}^{8}, \nonumber \\
        c_{0} &= -4.836845421239541, \nonumber \\
        c_{2} &= +8.400000000000002, \nonumber \\
        c_{4} &= -5.626807894353868, \nonumber \\
        c_{6} &= +1.7100830984547917, \nonumber \\
        c_{8} &= -0.19798989873223335.
    \end{align}

    \subsubsection*{Software}
    We use LAMMPS~\cite{plimpton1995fast, lammps} to perform simulations of
    glass forming liquids and utilise the built-in conjugate-gradient minimiser
    with a force tolerance $\sqrt{\sum_{i=1}^{N} {\lvert F_{i} \rvert}^{2}}
    \text{ of } 10^{-11}$ to reach inherent structures from an equilibrium
    sample at a temperature $T=1.0$ (lj units).  Eigenvalue calculations are
    performed using the \textit{dsyevr} LAPACK~\cite{laug} routine in Intel
    MKL~\cite{intelmkl}.  Analyses are performed using
    Mathematica~\cite{mathematica} and SciPy~\cite{scipy}. Plotting is
    performed using Matplotlib~\cite{hunter2007matplotlib, matplotlib}.

\end{appendix}

\newpage
\end{widetext}

\end{document}